\documentclass{article}

\usepackage{amsmath,amsthm,amssymb,mathtools} 
\usepackage[affil-it]{authblk} 
\usepackage{color}    
\usepackage{etoolbox}
\usepackage{float} 
\usepackage[margin=2cm]{geometry} 
\usepackage[acronym,toc,shortcuts]{glossaries}
\usepackage{graphics} 
\usepackage{ifplatform} 
\usepackage{import}
\usepackage[pagewise,switch]{lineno} 
\usepackage{mathptmx} 
\usepackage{natbib}
\usepackage{relsize} 
\usepackage[section]{placeins} 
\usepackage{siunitx}  
\usepackage{subfigure}
\usepackage{times}   
\usepackage{units,xspace}
\usepackage{verbatim} 
\usepackage{xifthen}  

\newglossaryentry{oesophagus}{
	name={{\oe}sophagus},
	plural={{\oe}sophagi},
	description={canal from mouth to stomach}
}
\newglossaryentry{emptyset}{
	name={empty set},
	symbol={\ensuremath{\emptyset}},
	description={the set containing no elements}
}
\newacronym
  [longplural={diagonal matrices}]
  {dm} {DM}{diagonal matrix}
\makeglossaries

\newacronym[longplural={circular variances}]{cv}{CV}{circular variance}
\newacronym[longplural={cycles per degree}]{cpd}{CPD}{cycles per degree}
\newacronym{dog}{DOG}{difference of Gaussian}
\newacronym{gaba}{GABA}{gamma-aminobutyric acid}
\newacronym{lgn}{LGN}{lateral geniculate nucleus}
\newacronym{ntwk1}{FF}{feed forward}
\newacronym{ntwk2}{LI}{lateral inhibitory}
\newacronym{os}{OS}{orientation selectivity}
\newacronym{psp}{PSP}{post-synaptic potential}
\newacronym{pseudorf}{pseudo-RF}{pseudo receptive field}
\newacronym{retina}{retina}{retina}
\newacronym{rgc}{RGC}{retinal ganglion cells}
\newacronym{rf}{RF}{receptive field}
\newacronym{stdp}{STDP}{spike-timing dependent plasticity}
\newacronym{v1}{V1}{primary visual cortex}
\newacronym{wgn}{WGN}{white, Gaussian noise}

\DeclareMathSizes{7}{7}{5}{5} 

\newcommand{\processnext}[1]{%
  \ifx\listfinish#1\empty\else\listact{#1}\expandafter\processnext\fi}

\newcommand{\mediumspace}{\ensuremath{\:}}

\newcommand{\largespace}{\ensuremath{\quad}}
\newcommand{\verylargespace}{\ensuremath{\qquad}}

\ifx\isEmbedded\undefined

\else

\fi
\newcommand{\app}[1]{Appendix~#1}

\newcommand{\eq}[1]{Eq.~(#1)}

\newcommand{\fig}[1]{Fig.~#1}

\newcommand{\figpanel}[1]{\textbf{(#1)}}

\renewcommand{\sec}[1]{Section~#1}

\makeatletter

\newcommand{\Rmnum}[1]{\expandafter\@slowromancap\romannumeral #1@}
\makeatother

\newcommand{\imag}{{i\mkern1mu}}

\DeclareMathAlphabet{\mathpzc}{OT1}{pzc}{m}{it} 
\DeclareMathAlphabet{\mathcal}{OMS}{cmsy}{m}{n} 

\DeclarePairedDelimiter{\floor}{\lfloor}{\rfloor}
\DeclarePairedDelimiter{\ceil}{\lceil}{\rceil}
\DeclarePairedDelimiter\abs{\lvert}{\rvert}%
\DeclarePairedDelimiter\norm{\lVert}{\rVert}%
\makeatletter
\let\oldfloor\floor
\def\floor{\@ifstar{\oldfloor}{\oldfloor*}}

\makeatletter
\let\oldceil\ceil
\def\ceil{\@ifstar{\oldceil}{\oldceil*}}

\makeatletter
\let\oldabs\abs
\def\abs{\@ifstar{\oldabs}{\oldabs*}}

\let\oldnorm\norm
\def\norm{\@ifstar{\oldnorm}{\oldnorm*}}
\makeatother

\newcommand*\xoverline[2][0.75]{%
  \sbox{\myboxA}{$\math#2$}%
    \setbox\myboxB\null
    \ht\myboxB=\ht\myboxA%
	\dp\myboxB=\dp\myboxA%
	\wd\myboxB=#1\wd\myboxA
	\sbox\myboxB{$\math\overline{\copy\myboxB}$}
	\setlength\mylenA{\the\wd\myboxA}
	\addtolength\mylenA{-\the\wd\myboxB}%
	\ifdim\wd\myboxB<\wd\myboxA%
	  \rlap{\hskip 0.5\mylenA\usebox\myboxB}{\usebox\myboxA}%
    \else
	  \hskip -0.5\mylenA\rlap{\usebox\myboxA}{\hskip
	  		  0.5\mylenA\usebox\myboxB}%
    \fi
}

\newcommand{\avg}[1]{\ensuremath{\mean{#1}}}
\newcommand{\conv}{\ensuremath{\star}}
\newcommand{\conj}[1]{\ensuremath{\left(#1\right)^*}}

\renewcommand{\exp}[1]{\mathrm{exp}\ensuremath{\left(#1\right)}}

\newcommand{\ii}{\mathrm{i}}
\newcommand{\mean}[1]{\ensuremath{\overline{#1}}}

\newcommand{\vect}[1]{{\bf #1}}
\newcommand{\icol}[2]{
   \left( #1, #2 \right)
}

\newcommand{\delsym}{\ensuremath{\Delta}}
\newcommand{\delay}[4] 
{ \ifthenelse{\isempty{#1}}%
   {\ensuremath{\delsym}}
   {\ifthenelse{\isempty{#3}}%
	  {\ensuremath{\delsym_{#1#2}}}
      {\ensuremath{\delsym_{#1#2}^{#3#4}}}
   }
}

\newcommand{\cxcoeff}{\ensuremath{A}}

\newcommand{\distancesym}{\ensuremath{d}}
\newcommand{\distance}[4] 
{ \ifthenelse{\isempty{#3}}
	{\ensuremath{\distancesym_{#1#2}}\xspace}
    {\ensuremath{\distancesym_{#1#2}^{#4}}\xspace}
}

\newcommand{\distlayerSqu}[2]{
{ \ifthenelse{\isempty{#1}}
	{\ensuremath{\distancesym^2}}
    {\ensuremath{(\distancesym^{#1#2})^2}}
}}	
\newcommand{\expect}[1]{\ensuremath{\mathrm{E}\Bigl[#1\Bigr]}}
\newcommand{\posTwoD}[2]{\ensuremath{\vect{x}_{#1#2}}}
\newcommand{\poscont}{\ensuremath{\vect{x}}} 
\newcommand{\posr}{\ensuremath{r}}
\newcommand{\posri}[2]{\ensuremath{r_{#1#2}}} 

\newcommand{\posth}{\ensuremath{\theta}}
\newcommand{\posthi}[2]{\ensuremath{\theta_{#1#2}}} 

\newcommand{\postwocont}{\ensuremath{\vect{x}'}} 
\newcommand{\posintx}{\ensuremath{\tilde{x}}} 
\newcommand{\posintr}{\ensuremath{\tilde{r}}}

\newcommand{\postwoth}{\ensuremath{\tilde{\posth}}}

\newcommand{\posvec}[2]{\ensuremath{\left[#1,#2\right]}}
\newcommand{\posx}[2]{\ensuremath{x_{#1#2}}}
\newcommand{\posy}[2]{\ensuremath{y_{#1#2}}}
\newcommand{\posxcont}{\ensuremath{x}} 
\newcommand{\posycont}{\ensuremath{y}} 
\newcommand{\posxtwocont}{\ensuremath{\tilde{x}}} 
\newcommand{\posytwocont}{\ensuremath{\tilde{y}}} 

\newcommand{\atDist}[1]{\ensuremath{{#1}\left(\distance{}{}{}{}\right)}} 
 
\newcommand{\timesym}{\ensuremath{t}} 
\newcommand{\atTime}[1]{\ensuremath{#1\left( \timesym \right)}}

\newcommand{\filt}{\ensuremath{\psp}}

\newcommand{\psp}{\ensuremath{\epsilon}}

\newcommand{\scov}[2]{\ensuremath{\mathrm{cov}\left(#1, \, #2\right)}}

\newcommand{\taul}[2]{\ensuremath{\tau_l}} 
\newcommand{\taur}[2]{\ensuremath{\tau_r}} 

\newcommand{\bessel}[2]{\ensuremath{I_{#1}\left(#2\right)}}

\newcommand{\centreN}{\ensuremath{c}}

\newcommand{\ensemble}[1]{\ensuremath{\left\langle #1 \right\rangle}}
\newcommand{\etalayer}[1]{\ensuremath{\eta_{#1}}}
\newcommand{\expsum}{\ensuremath{n}}
\newcommand{\expsumtwo}{\ensuremath{n'}}

\newcommand{\fouriersumint}{\ensuremath{l'}}
\newcommand{\fouriersum}{\ensuremath{l}}
\newcommand{\cossum}{\ensuremath{k}}
\newcommand{\coscoeff}[2]{\ensuremath{g_{#1}\left(#2\right)}}
\newcommand{\sincoeff}[2]{\ensuremath{\tilde{g}_{#1}\left(#2\right)}}
\newcommand{\cosbr}[2]  
{ \ifthenelse{\isempty{#2}}%
	{\ensuremath{\cos\left(#1\right)}}
	{\ensuremath{\cos^{#2}\left(#1\right)}}
}
\newcommand{\sinbr}[1]{\ensuremath{\sin\left(#1\right)}} 

\newcommand{\eigdecay}{\ensuremath{C}} 
\newcommand{\eigAllCoeff}{\ensuremath{\Lambda_0}} 
\newcommand{\eignorm}[2]{\ensuremath{N_{#1,#2}}}  
\newcommand{\eigxindex}{\ensuremath{u}}
\newcommand{\eigyindex}{\ensuremath{v}}
\newcommand{\eigLindex}{\ensuremath{n}} 
\newcommand{\eigvec}[2]{\ensuremath{\vect{v}_{#1}\left(#2\right)}}

\newcommand{\eigvecSimple}[2]{\ensuremath{\vect{v}_{#1}^0\left(#2\right)}}
\newcommand{\eigvecPerturb}[2]{\ensuremath{\vect{v}_{#1}^1\left(#2\right)}}
\newcommand{\eigvecFull}[2]{\ensuremath{\vect{v}_{#1}\left(#2\right)}}
\newcommand{\eigvalueSimple}[2]{\ensuremath{\lambda_{#1,#2}^0}}
\newcommand{\eigvaluePerturb}[2]{\ensuremath{\lambda_{#1,#2}^1}}
\newcommand{\eigvalueFull}[2]{\ensuremath{\lambda_{#1,#2}}}
\newcommand{\eigvalue}[2]{
{ \ifthenelse{\isempty{#1}}
	{\ensuremath{\lambda}}
    {\ifthenelse{\isempty{#2}}
		{\ensuremath{\lambda_{#1}}}
		{\ensuremath{\lambda_{#1,#2}}}}
}}

\newcommand{\fo}[1]
{ \ifthenelse{\isempty{#1}}%
	{\ensuremath{f_{0}^2}}
	{\ifthenelse{\equal{\detokenize{#1}}{\detokenize{-}}}
		{\ensuremath{f_{0}^{-2}}}
        {\ensuremath{f_{0,#1}^2}}
	}
}
\newcommand{\expcoeff}{\ensuremath{\alpha}}

\newcommand{\operatorSimple}{\ensuremath{\mathrm{H}^0}}
\newcommand{\operatorPerturb}{\ensuremath{\mathrm{H}^1}}
\newcommand{\operatorFull}{\ensuremath{\mathrm{H}}}

\newcommand{\hermitecoeff}[1]{\ensuremath{\frac{1}{\sqrt{2^{#1}#1!}}}} 
\newcommand{\hermite}[2]{\ensuremath{\mathrm{H}_{#1}\left(#2\right)}} 

\newcommand{\laguerre}[3]
{ \ifthenelse{\isempty{#3}}
	{\mathrm{L}\ensuremath{_{#1}^{#2}}}
	{\mathrm{L}\ensuremath{_{#1}^{#2}\left(#3\right)}} 
}
\newcommand{\inputlayer}{\ensuremath{\layer{A}}}

\newcommand{\kone}[2]{\ensuremath{k_1^{#1#2}}}
\newcommand{\ktwo}[2]{\ensuremath{k_2^{#1#2}}}
\newcommand{\layerrate}[1]{\ensuremath{\lambda^{#1}}}

\newcommand{\layerstd}[1]{\ensuremath{\sigma^{#1}}}
\newcommand{\layer}[1]{\textit{#1}} 
\newcommand{\Lcov}[3]{\ensuremath{Q_{#2#3}^{#1}}} 
\newcommand{\lcov}[2] 
{ \ifthenelse{\isempty{#1}}
	{\ensuremath{Q}}
	{\ensuremath{Q\left({#1,#2}\right)}}
}
\newcommand{\Lrate}[3]  
{ \ifthenelse{\isempty{#3}}
	{ \ensuremath{f_{#1}^{#2}}\xspace}
	{ \ifthenelse{\equal{ \detokenize{#3} }{ \detokenize{\filt} } } 
	    { \ensuremath{ ( f_{#1}^{#2} \conv \filt ) } }
		{ \ensuremath{ ( f_{#1}^{#2(\delsym)} \conv \filt ) } }
	}
}

\newcommand{\Lwgtsym}{\ensuremath{w}}
\newcommand{\Lweight}[4]{\ensuremath{{\Lwgtsym}^{#1#2}_{#3#4}}} 
\newcommand{\Lwchange}[4]{\ensuremath{\dot{\Lwgtsym}^{#1#2}_{#3#4}}} 
\newcommand{\LwgtMin}[2]{\ensuremath{\Lwgtsym_{\textrm{min}}^{#1#2}}} 
\newcommand{\LwgtMax}[2]{\ensuremath{\Lwgtsym_{\textrm{max}}^{#1#2}}} 
\newcommand{\Lwgtcont}[2]{ 
{\ifthenelse{{\isempty{#2}}}
	{\ensuremath{\Lwgtsym\left({#1}\right)}}
	{\ensuremath{\Lwgtsym\left({#1,#2}\right)}}
}}
\newcommand{\Lwderiv}[2]{
{\ifthenelse{{\isempty{#2}}}
	{\ensuremath{\dot{\mathlarger{v}}\left(#1\right)}}
	{\ensuremath{\dot{\mathlarger{v}}\left(#1,#2\right)}}
}}

\newcommand{\meanLweight}[2]{\ensuremath{\avg{\Lwgtsym}^{#1#2}}} 
\newcommand{\meanrate}[1]{\ensuremath{\mean{\Lrate{}{#1}{}}}}

\newcommand{\N}[2]{\ensuremath{N^{#1#2}}} 

\newcommand{\Nprepost}[4]{\ensuremath
{\ifthenelse{{\isempty{#3}}}
	{\ensuremath{N_{#1#2}}}
	{\ensuremath{N_{#1#2}^{#3#4}}}
}} 
\newcommand{\Nshared}[2] 
{\ifthenelse{{\isempty{#2}}}
	{\ensuremath{N^{#1#1}}}
	{\ensuremath{N^{#1#2}}}
}

\newcommand{\preN}{\ensuremath{m}}
\newcommand{\preNtwo}{\ensuremath{n}}
\newcommand{\postN}{\ensuremath{i}}
\newcommand{\postNtwo}{\ensuremath{j}}
\newcommand{\postpostN}{\ensuremath{p}}

\newcommand{\radsym}{\ensuremath{\sigma}}
\newcommand{\radius}[2]{\ensuremath{\radsym^{#1#2}}}
\newcommand{\radiusvar}[2]{\ensuremath{(\radsym^{#1#2})^2}}

\newcommand{\Ra}[1]
{ \ifthenelse{\isempty{#1}}%
	{\ensuremath{R_a}}
    {\ensuremath{R_a^{#1}}}
}
\newcommand{\Rb}[2]{ }
\newcommand{\RbSq}[2]{ }

\newcommand{\Sum}[3]{\ensuremath{\sum\limits_{#1=#2}^{#3}}}

\DeclareMathAlphabet{\mathpzc}{OT1}{pzc}{m}{it} 

\newcommand{\tab}{\hspace{5mm}}

\newcommand{\dist}[2]{\ensuremath{\distfont{#1}\left(#2\right)}}
\newcommand{\distfont}[1]
{ \ifthenelse{\isempty{#1}}%
	{\mathscr{#1}}
	{\ifthenelse{\equal{#1}{t}}
		{\ensuremath{\mathlarger{\mathlarger{\mathpzc{#1}}}}}
		{\ifthenelse{\equal{#1}{Lapl}}
			{\ensuremath{\mathlarger{\mathlarger{\mathpzc{#1}}}}}
			{\ifthenelse{\equal{#1}{Poisson}}
			   {\ensuremath{\mathlarger{\mathpzc{Poisson}}}}
			   {\ensuremath{\mathpzc{#1}}}
			}
		}
	}
}

\newcommand{\probdist}[3]
{ \ifthenelse{\isempty{#3}}%
	{p_{#1}\left(#2\right)}
    {p_{#1}\left(#2\,;\,#3\right)}
}
\newcommand{\fnin}[1]
{ \ifthenelse{\isempty{#1}}%
	{}
	{\ensuremath{_{#1}}}
}
\newcommand{\sfnin}[1]
{ \ifthenelse{\isempty{#1}}%
	{}
	{\ensuremath{\left({#1}\right)}}
}
\newcommand{\ssym}[1]
{ \ifthenelse{\isempty{#1}}%
	{}
	{\ensuremath{\mathrm{#1}}}
}
\newcommand{\psym}[1]
{ \ifthenelse{\isempty{#1}}%
	{}
	{\ensuremath{\hat{#1}}}
}

\bibpunct{(}{)}{;}{a}{,}{,} 

\makeatletter

\newcommand{\isEmbedded}{true}

\patchcmd{\NAT@test}{\else \NAT@nm}{\else \NAT@nmfmt{\NAT@nm}}{}{}

\DeclareRobustCommand\citepos
  {\begingroup
   \let\NAT@nmfmt\NAT@posfmt
   \NAT@swafalse\let\NAT@ctype\z@\NAT@partrue
   \@ifstar{\NAT@fulltrue\NAT@citetp}{\NAT@fullfalse\NAT@citetp}}

\let\NAT@orig@nmfmt\NAT@nmfmt
\def\NAT@posfmt#1{\NAT@orig@nmfmt{#1's}}

\begin{document}


\title{Emergence of radial orientation selectivity: Effect of cell density changes and eccentricity in a layered network}
\author[a]{Catherine E. Davey}
\author[a]{David B. Grayden}
\author[a]{Anthony N. Burkitt}
\affil[a]{Department of Biomedical Engineering, The University of Melbourne, VIC 3010, Australia}

\maketitle



\section*{Abstract}

An account of how simple cells can emerge in the absence of structured environmental 
input via a self-organised learning process has been provided by Linsker (1986); 
this work empirically showed the emergence of spatial-opponent cells as a result 
of structure in the initial synaptic connectivity distribution when the visual 
system is driven entirely by input noise. In this paper, the complete set of 
eigenfunctions and eigenvalues for a three-layer network is analytically derived 
for the first time. As a first step, a simplified learning equation is considered 
for which the homeostatic parameters are set to zero. This is then extended to an 
analysis of the eigenfunctions of the full learning equation, including non-zero 
homeostatic parameters, using a perturbation analysis. These results extend the 
previous analysis of the  Linsker (1986) network to allow for radially dependent 
cell density, as found in the retina. The results establish that radially biased 
orientation selectivity emerges in the third layer when cell density in the first 
layer changes with eccentricity; i.e., distance to the centre of the lamina. 
This provides a potential mechanism for the emergence of radial orientation in the 
primary visual cortex before eye opening and the onset of structured visual input 
after birth. 

\section{Introduction}\label{sec:intro}
   
Synaptic plasticity underpins our understanding of learning in neural systems 
as it is the mechanism that describes how synaptic 
weights change in response to sensory inputs. Plasticity has traditionally 
been modelled as rate based, in which synaptic weights change in response to 
short-time averaged pre- and postsynaptic neuron spiking rates. Over the past 
two decades, the importance of pre- and postsynaptic neuron spike timing has been 
recognized, particularly for contexts in which high-resolution temporal information 
is involved at microsecond resolution \citep{KemGerHem99, GerKemvanWag96}, prompting 
the emergence of \gls{stdp} \citep{GerKemvanWag96,MarLubFroSak97}. Spike-based 
plasticity updates synaptic strength in response to the relative timing of pre- 
and postsynaptic spikes, amplifying synaptic strength if the presynaptic neuron 
contributes to the postsynaptic neuron's spike, and depressing a synapse 
if the presynaptic neuron fires after the postsynaptic neuron and thus did not 
contribute to its spike. 

Plasticity mechanisms have played a fundamental role in explaining the 
emergence of simple cells in the early layers of cortical processing, such as 
the \gls{v1}. Plasticity has successfully explained the emergence of simple cells 
such as orientation selective cells \citep{BieCooMun82,WimGerHem98,Yam02}, 
direction selective cells \citep{WimWenMilvanH97a,WimWenMilvanH97b,SenBuc03}, 
ocular dominance \citep{Mill90}, and feature maps in which sensitivity to a 
particular feature changes as the layer is traversed \citep{Goo007}. 

Much of the research on learning in cortical networks has been empirical and 
computational because of the analytical complexity of learning in response to 
parameters that describe the number of layers, connectivity structure, and neuron type. 
A notable exception is the analysis of the network proposed by \citet{Lin86a} in 
which  the emergence of a spatial opponent cell in the third layer of a three-layer 
network of Poisson neurons with Gaussian connectivity kernels was described. Learning 
in this network is a linear function of correlation in presynaptic neural activity, with 
two learning constants that control the homeostatic equilibrium. The linearity 
of the learning system enables an eigenfunction analysis to be used to identify 
the independent contributors to a postsynaptic neuron's synaptic weight structure. 
Eigenvalues provide a way to distinguish the eigenfunctions that are the most 
significant contributors, and hence determine the receptive field of the postsynaptic 
neuron. 

Although \citet{Lin86a} focused on empirical results, there has been 
significant work aimed at extending the analytical framework for the network that he 
proposed. \citet{MacMil90} proposed the first three radial eigenfunctions based 
on the work by \citet{Tang90}, but without providing a derivation. The proposed 
eigenfunctions were for a simplified learning system in which homeostatic constants 
were assumed zero so that all plasticity was driven by correlation between 
presynaptic inputs and there was no non-competitive plasticity. They provided 
an empirical examination of the impact of non-zero homeostatic constants, showing 
that the eigenfunction of the leading eigenvalue can change in response to a 
change in the homeostatic equilibrium. 

\citet{Mill90} employed \citepos{Lin86a} network in a model of learning in the 
primary visual cortex, with overlapping left and right eye inputs processed by 
the \gls{lgn}. The network structure prompted correlation and anti-correlation 
in two afferents originating from either the same eye or the opposite eye, 
leading to the emergence of an ocular dominance feature map. \citet{Mill90} 
provided a description of an analytical derivation for the eigenfunctions of 
ocular dominance feature maps across the cortex.

\citet{WimGerHem98} extended \citepos{Lin86a} network by incorporating lateral 
inhibitory connections in the third layer, showing the emergence of orientation 
selective cells in the third layer. They provided a derivation of Cartesian 
eigenfunctions for learning with homeostatic constants set to zero and empirically 
extended the solution to the general learning equation with non-zero homeostatic 
constants. They simulated the development of an orientation selective feature map 
distributed across the primary visual cortex using a model slightly more complex 
than that for which they derived the eigenfunctions. 

Analytical solutions to \citepos{Lin86a} learning system have played a central
role in explaining the emergence of spatial opponent and orientation selective 
cells in the network. However, thus far, no general analytical solution has been 
provided, with analytical results to date being for the simplified system in which the 
homeostatic constants are set to zero. We provide here a solution for the eigenfunctions 
of \citepos{Lin86a} network in polar coordinates. As the system is radially 
symmetric, polar coordinates provide a natural coordinate system that enables an 
straightforward extension of polar eigenfunctions to the general learning system with non-zero 
homeostatic constants. One of the benefits of a full analytical solution for the 
network is insight into why the receptive field changes in response to changes 
in the homeostatic equilibrium and the framework to determine exactly when this 
change occurs. 

Thus far, the original network proposed by \citet{Lin86a} and used in the 
subsequent analytical analyses of \citet{Mill90} and \citet{WimGerHem98} made 
an assumption that cells within each layer were evenly distributed and that 
receptive fields of all cells in a layer were statistically identical; i.e., drawn 
from the same synaptic connectivity distribution. To date, there has been no 
exploration of the impact of relaxing this assumption. However, it is known  
that some biological cell layers show an uneven density of cells across 
the lamina and contain receptive fields with different statistical properties. 
For example, the retina is well known to have cell density changes 
as a function of eccentricity \citep{SjoOlsPopCon99,Wat14} and receptive field 
sizes of neurons in the primary visual cortex increase with stimulus eccentricity 
\citep{WurMinYaz13,SmiSinWilGre01}. Furthermore, it is well established that 
orientation selectivity in the primary visual cortex is biased towards radial 
orientation in that an orientation selective neuron in the primary visual 
cortex is more likely to be oriented towards the centre of the retina 
\citep{RodRevPig04}. In this study, we explore the impact of radially dependent 
synaptic connection distributions on emerging receptive field properties in the 
third layer of \citepos{Lin86a} network and show how introducing radially dependent 
synaptic connectivity distributions in the first layer results in the emergence of 
radial orientation selectivity in the third layer of the network.

This paper is organised as follows. \sec{\ref{sec:ntwk}} introduces the network
and neuron models used, based on \citepos{Lin86a} network. Radial eigenfunctions 
and eigenvalues are analytically derived for the simplified learning equation, 
for which the homeostatic parameters are set to zero, and then extended via 
perturbation analysis to the full system in \sec{\ref{sec:radial_eigen}}.
Eigenfunctions and eigenvalues are also derived in Cartesian coordinates in
\sec{\ref{sec:radial_eigen}} and compared to the radial eigenfunctions.
Finally, we show in \sec{\ref{sec:emergence_radial}} that the introduction of 
radially dependent synaptic connectivity distributions in the first layer 
generates radial orientation selectivity in the third layer of the network.

\section{Methods}
    \subsection{Network specification}\label{sec:ntwk}
   
\begin{figure}[ht!]
\centering
	\includegraphics[width=0.4\textwidth]{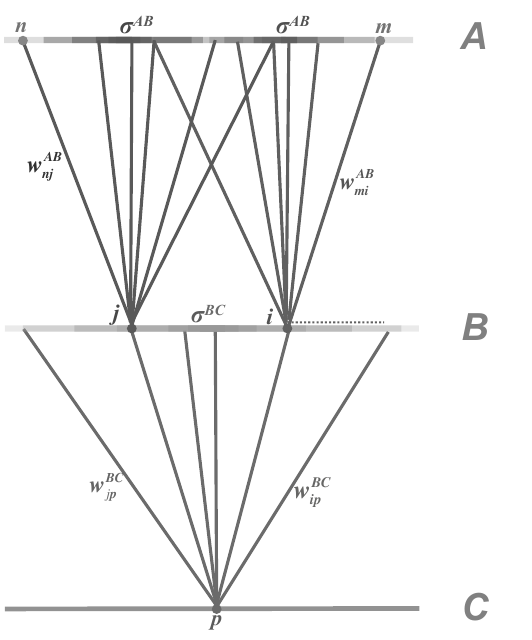}
	\caption{Schematic diagram of the three layered feed-forward network.  
		Layer $\layer{A}$ neurons, $\preN$ and $\preNtwo$, feed into 
		layer $\layer{B}$ neurons, $\postN$ and $\postNtwo$, respectively, 
		which in turn feed to a layer $\layer{C}$ neuron, $\postpostN$ (green). 
		Synaptic connections between neurons are shown as solid, coloured lines 
		connecting from a presynaptic neuron to a postsynaptic neuron of the same 
		colour. The synaptic strength, for example between neurons $\preN$ and $\postN$, 
		is denoted $\Lweight{\layer{A}}{\layer{B}}{\preN}{\postN}$. Synaptic connection  
		distributions are homogeneous within a layer and modelled as being Gaussian, 
		parameterised by distance-dependent standard deviation (radius) denoted 
		$\radius{\layer{A}}{\layer{B}}$ between layers $\layer{A}$ and $\layer{B}$ 
		and $\radius{\layer{B}}{\layer{C}}$ between layers $\layer{B}$ and $\layer{C}$. 
		The probability of a postsynaptic neuron having a presynaptic connection from
		a particular position in the layer is depicted by the intensity of the colour 
		of the presynaptic layer. Not shown in the diagram is the expected 
		number of presynaptic connections input to a neuron, denoted by 
		$\N{\layer{A}}{\layer{B}}$ and $\N{\layer{B}}{\layer{C}}$ for postsynaptic 
		neurons in layers $\layer{B}$ and $\layer{C}$, respectively. The radial 
		distance between two neurons within a lamina, for example between neuron $\preN$ 
		from layer $\layer{A}$ and $\postN$ from layer $\layer{B}$, is denoted 
		$d_{\preN\postN}^{\layer{B}}$. 
		\label{fig:network}}
\end{figure}

Following \citet{Lin86a}, we consider a three-layer, feed-forward topographical 
network. The network is driven by spontaneous neural activity in the first layer, 
layer $\layer{A}$, which inputs to layer $\layer{B}$, which in turn inputs to 
layer $\layer{C}$, as shown schematically in \fig{\ref{fig:network}}. Layers  
comprise populations of homogeneous neurons, equispaced in a square grid across 
the layer. The distance between the parallel layers is assumed to dominate sufficiently 
such that propagation delay experienced by action potentials from the presynaptic 
layer can be assumed approximately equal. Neurons $\preN$ and $\preNtwo$ of layer 
$\layer{A}$ have synaptic inputs to neurons $\postN$ and $\postNtwo$ of layer 
$\layer{B}$, respectively, which both input to neuron $\postpostN$ of layer $\layer{C}$. 

Each postsynaptic neuron has a Gaussian synaptic connection 
distribution, centred on its two-dimensional position in the lamina, which ensures 
that radially proximate neurons are more likely to connect to it than a neuron more 
distal in the presynaptic lamina. The connectivity distributions are parameterised 
by a standard deviation (radius) that is homogeneous across a layer, denoted 
$\radius{\layer{A}}{\layer{B}}$ and $\radius{\layer{B}}{\layer{C}}$, 
for synaptic connections between layers $\layer{A}$ and $\layer{B}$ and layers 
$\layer{B}$ and $\layer{C}$, respectively. Consequently, the probability of 
neuron $\preN$ in layer $\layer{A}$ connecting to neuron $\postN$ in layer 
$\layer{B}$ is given by 
\begin{align}\label{eq:2DGauss}
	\probdist{N}{ \icol{ \posx{\preN}{\postN} }{ \posy{\preN}{\postN}} }
				{ \frac{1}{2} \radiusvar{\layer{A}}{\layer{B}} }
  &= 
	\frac{1}{\pi\radiusvar{\layer{A}}{\layer{B}}} 
	\exp{- \frac{ \posx{\preN}{\postN}^2 
				+ \posy{\preN}{\postN}^2 }
				{ \radiusvar{\layer{A}}{\layer{B}}} } \, ,
\end{align}
where $\icol{\posx{\preN}{\postN} }{ \posy{\preN}{\postN}}$ is the two-dimensional 
radial distance between $\preN$ and $\postN$. Note that this definition differs 
from the standard definition by a factor of $\sqrt{2}$ in accordance with the 
definition used by \citet{Lin86a}, and is specifically chosen for later convenience. 

For postsynaptic neurons in layer $\layer{C}$, it is useful to write the connection 
probability in polar coordinates by assuming, without loss of generality, that 
the postsynaptic neuron is at position $\icol{0}{0}$. The probability of presynaptic 
neuron $\postNtwo$ in layer $\layer{B}$ connecting to postsynaptic neuron $\postpostN$ 
in layer $\layer{C}$ in polar coordinates is then
\begin{align}\label{eq:2DGauss_polar}
	\probdist{N}{ \icol{\posri{\postNtwo}{\postpostN} }{ \posthi{\postNtwo}{\postpostN}} }
				{ \frac{1}{ \radiusvar{\layer{B}}{\layer{C}} } }
  &= 
	\frac{1}{ \pi \radiusvar{\layer{B}}{\layer{C}} } 
	\exp{- \frac{ \posri{\postNtwo}{\postpostN}^2 }
				{ \radiusvar{\layer{B}}{\layer{C}}} },
\end{align}
where $\posri{\postNtwo}{\postpostN}$ is the radial distance from the centre of 
the lamina to neuron $\postN$ in layer $\layer{}C$, and 
$\posthi{\postNtwo}{\postpostN}$ is the angle to $\postN$ within the two-dimensional 
lamina.

\citet{Lin86a} showed that the Gaussian connectivity distributions introduce spatial 
correlations in the inputs to layer $\layer{B}$ neurons despite spontaneous neural 
activity in layer $\layer{A}$ being uncorrelated. Layer $\layer{B}$ neurons 
that are spatially more proximate will have a greater number of shared connections, 
and therefore more correlated input, when compared to layer $\layer{B}$ neurons 
that are positioned further apart in the lamina. The expected number of shared 
presynaptic inputs between two postsynaptic neurons in layer $\layer{B}$ is shown 
to be (see \app{\ref{app:sharedInputs}} for full derivation)
\begin{align}\label{eq:sharedInputs}
   \expect{\atDist{\Nshared{\layer{B}}{}}}
 &=
   \frac{ (\N{\layer{A}}{\layer{B}})^2 }{ 2\pi\radiusvar{\layer{A}}{\layer{B}} }
   \exp{ -\frac{ (\distance{\postN}{\postNtwo}{\layer{B}}{\layer{B}})^2 }
			   {2\radiusvar{\layer{A}}{\layer{B}}}} \, ,
\end{align}
where $\N{\layer{A}}{\layer{B}}$ denotes the expected number of synaptic connections 
from layer $\layer{A}$ to each neuron in layer $\layer{B}$, and 
$\distance{\postN}{\postNtwo}{\layer{B}}{\layer{B}}$ represents the distance 
between neurons $\postN$ and $\postNtwo$ such that 
$  \distance{\postN}{\postNtwo}{\layer{B}}{\layer{B}} 
 = \sqrt{\posx{\preN}{\postN}^2 + \posy{\preN}{\postN}^2}$. 

	\subsection{Neuron model}\label{sec:neuronModel}
	   
The network is driven by spontaneous Poisson activity of the layer $\layer{A}$ 
neurons. This implies that there are no spike-based temporal correlations between 
input and output neurons other than what is captured in the rate-based signals 
and that the rates change slowly when compared to the period over which they are averaged 
 \citep{KemGervan99}. Activity of a layer $\layer{A}$ neuron is modelled as 
$\atTime{ \Lrate{\preN}{\layer{A}}{} } \sim \dist{Poisson}{\layerrate{\layer{A}}}$,
where $\atTime{ \Lrate{\preN}{\layer{A}}{} }$ is the spiking rate of layer 
$\layer{A}$ neuron $\preN$ at time $\timesym$. 

As in \citet{Lin86a}, we use a Poisson neuron model so that the network is linear 
when operating within the weight bounds, discussed below. The update equations for 
neural activity in layers $\layer{B}$ and $\layer{C}$ are
\begin{subequations}\label{eq:layer_rate}
\begin{align}
  \atTime{ \Lrate{\postN}{\layer{B}}{} }
     &= \Ra{\layer{B}} + \Rb{\layer{\layer{A}}}{\layer{B}}
	    \sum\limits_{\preN} \atTime{\Lweight{\layer{A}}{\layer{B}}{\preN}{\postN}}
		                    \atTime{\Lrate{\preN}{\layer{A}}{}},
  \label{eq:layerB_rate}  \\
  \atTime{ \Lrate{\postpostN}{\layer{C}}{} }
     &= \Ra{\layer{C}} + \Rb{\layer{\layer{B}}}{\layer{C}}
	    \sum\limits_{\postN} \atTime{\Lweight{\layer{B}}{\layer{C}}{\postN}{\postpostN}}
		                     \atTime{\Lrate{\postN}{\layer{B}}{}},
  \label{eq:layerC_rate}
\end{align}
\end{subequations}
where $\Ra{ \layer{B} }$, $\Ra{ \layer{C} }$ denote spontaneous firing rates, and  
$\atTime{ \Lweight{ \layer{A} }{ \layer{B} }{ \preN}{\postN} }$, 
$\atTime{ \Lweight{ \layer{B} }{ \layer{C} }{ \postN }{ \postpostN} }$ depict 
synaptic strengths between neurons $\preN$ and $\postN$ in layers $\layer{A}$ 
and $\layer{B}$, respectively, and neurons $\postN$ and $\postpostN$ in layers 
$\layer{B}$ and $\layer{C}$, respectively. 
Note that an implicit assumption in this Poisson model of neural activity is that 
propagation delay is negligible or, equivalently, is dominated by inter-layer 
distances between neurons and, therefore, can  be considered homogeneous across  
all inputs to a postsynaptic neuron.

	\subsection{Learning dynamics}\label{sec:learning} 
	   
The adiabatic approximation in neural learning is that incremental weight changes 
occur slowly with respect to neural dynamics, which occur on a millisecond timescale. 
Furthermore, neurons within the same population are assumed to have the same 
statistical properties of neural activity and synaptic connectivity. Consequently, 
the system is ergodic and the spike rate can be determined from the ensemble average 
or from a temporal mean over the timescale of learning. Under these assumptions, 
the learning equation can be expressed as a differential equation \citep{Lin86a}. 
The general learning equations for synaptic weights between neurons in layers 
$\layer{A}$ and $\layer{B}$ and synapses connecting layers $\layer{B}$ and 
$\layer{C}$ is given by \citep{Lin86a}
\begin{subequations}\label{eq:learningRule_Linsker}
\begin{align}
  \etalayer{} \Lwchange{\layer{A}}{\layer{B}}{\preN}{\postN} 
 &=
  \kone{\layer{A}}{\layer{B}} + \frac{1}{\N{\layer{A}}{\layer{B}}}
  \sum_{\preNtwo} \Lweight{\layer{A}}{\layer{B}}{\preNtwo}{\postN}
                  \left( \Lcov{\layer{A}}{\preN}{\preNtwo} 
				       + \ktwo{\layer{A}}{\layer{B}} 
				  \right) \, ,
 \verylargespace 
	\LwgtMin{}{} \leq \Lweight{\inputlayer}{\layer{B}}{\preN}{\postN} \leq \LwgtMax{}{} \, , 
 \label{eq:learningRule_Linsker_AB} \\
  \etalayer{} \Lwchange{\layer{A}}{\layer{B}}{\postN}{\postpostN} 
 &=
  \kone{\layer{B}}{\layer{C}} + \frac{1}{\N{\layer{B}}{\layer{C}}}
  \sum_{\postNtwo} \Lweight{\layer{B}}{\layer{C}}{\postNtwo}{\postpostN}
                  \left( \Lcov{\layer{B}}{\postN}{\postNtwo} 
				       + \ktwo{\layer{B}}{\layer{C}} 
				  \right) \, ,
 \verylargespace 
	\LwgtMin{}{} \leq \Lweight{\inputlayer}{\layer{B}}{\preN}{\postN} \leq \LwgtMax{}{} \, , 
 \label{eq:learningRule_Linsker_BC}
\end{align}
\end{subequations}
where $\etalayer{} \ll 1$ is the learning rate that ensures that learning is slow 
on a millisecond timescale, $\LwgtMin{}{}$ and $\LwgtMax{}{}$ are the lower and 
upper bounds on the weights, respectively, and the parameters 
$\kone{\layer{A}}{\layer{B}}$, $\ktwo{\layer{A}}{\layer{B}}$, 
$\kone{\layer{B}}{\layer{C}}$, $\ktwo{\layer{B}}{\layer{C}}$ 
are layer specific constants controlling homeostasis (i.e., independent of the
correlation structure of the inputs). The definition for normalised covariance 
has the same structure for each layer; for example, the normalised covariance 
$\Lcov{ \layer{A}}{\preN}{\preNtwo}$
between layer $\layer{A}$ neurons $\preN$ and $\preNtwo$ is defined by 
\begin{equation}
 \Lcov{ \layer{A}}{\preN}{\preNtwo} 
 = \frac{ \ensemble{\Lrate{\preN}{\layer{A}}{}    - \meanrate{\layer{A}}}
          \ensemble{\Lrate{\preNtwo}{\layer{A}}{} - \meanrate{\layer{A}}}}
          { \fo{} },
\label{eq:norm_covar}
\end{equation}
where $\ensemble{}$ depicts the ensemble average, $\meanrate{\layer{A}}$ denotes the 
temporal average of layer $\layer{A}$ neural activity, and $\fo{}$ is a scaling 
factor to normalise the covariance matrix $\Lcov{}{}{}$.

For a Gaussian synaptic density distribution, the covariance between layer 
$\layer{B}$ neurons is a function of the radial distance separating the neurons, 
as described in Appendix~\ref{app:covLayerB},
\begin{align}\label{eq:Linsker_cov}
   \scov{\Lrate{\preN}{\layer{B}}{}}{\Lrate{\preNtwo}{\layer{B}}{}}
 &= 
   \frac{ \left( \Rb{\layer{B}}{\layer{C}} \N{\layer{B}}{\layer{C}} \layerstd{\layer{A}} 
		  \right)^2 }
        { 2\pi\radiusvar{\layer{A}}{\layer{B}} }
   \exp{-\frac{\distance{\preN}{\preNtwo}{\layer{B}}{\layer{B}}}
	          {2\radiusvar{\layer{A}}{\layer{B}}}}. 
\end{align}
Note that vector distances are given with respect to the layer \layer{C} cell, 
so we define the two-dimensional radial distance of presynaptic 
neuron, $\preN$, from postsynaptic neuron, $\postN$, by vector, 
$\posTwoD{\preN}{\postN} = [\posx{\preN}{\postN},\posy{\preN}{\postN}]'$, and 
$ \distance{\postN}{\postNtwo}{\layer{B}}{\layer{B}} 
= \norm{\posTwoD{\preN}{\postN}-{\posTwoD{\preNtwo}{\postN}}}$.
Only radial distances are considered, so that distances between layers are 
assumed to have negligible impact on learning dynamics since the inter-layer
transmission delay is uniform.

Normalising this result and incorporating it into 
\eq{\ref{eq:learningRule_Linsker}} gives the learning equation
\begin{align}\label{eq:cdot_BCcov}
  \etalayer{} \Lwchange{}{}{\preN}{\postN} 
 &= \kone{}{} + \frac{1}{\N{}{}}
    \sum_{\preNtwo} \Lweight{}{}{\preNtwo}{\postN}
          \left( \exp{- \frac{\abs{\posTwoD{\preN}{\postN} 
					              - \posTwoD{\preNtwo}{\postN}}^2}
				             {2\radiusvar{\layer{A}}{\layer{B}}}} 
	           + \ktwo{}{} \right),
\end{align}
where it is assumed that the covariance is normalised and we have removed the layer 
superscripts for readability.

It is assumed that a deeper layer is not learned until after its presynaptic layer 
has converged to a stable weight structure, and hence layers are learned 
sequentially. This accords with the approach employed by \citep{Lin86a} and does 
not impact the final weight structure across the network. Consequently, synapses 
connecting layers $\layer{A}$ and $\layer{B}$ evolve to a stable structure before 
learning begins for synapses connecting layers $\layer{B}$ and $\layer{C}$. 

\citet{Lin86a} demonstrated that individual synapses are unstable and, for 
excitatory synapses, all or all-but-one necessarily reach the upper bound, 
$\LwgtMax{}{}$. 
However, under an assumption of weak covariance of the inputs \citep{MacMil90}, 
the mean weight of synapses input to a postsynaptic neuron is not necessarily 
unstable but rather controlled by homeostatic mecehanisms. For excitatory 
connections, the mean weight of a postsynaptic neuron's synapses will converge to 
\begin{align}
   \meanLweight{}{} 
 &= 
  -\frac{ \kone{}{} }{ \ktwo{}{} },
   \verylargespace \textrm{if} \hspace{0.2cm}
   \ktwo{}{} < 0, \hspace{0.2cm} \textrm{and} \hspace{0.2cm}
   0 < \frac{\kone{}{} }{\ktwo{}{} } <  1,
\end{align}
where the conditions on $\kone{}{}$ and $\ktwo{}{}$ are required to ensure that  
the mean synaptic weight does not diverge to the bounds. 
For all synapses to grow until they reach the upper bound, it is required that 
$\kone{}{} + \ktwo{}{} > 0$. In this case, the system is unstable so that the 
mean synaptic weight grows until all individual synapses, or all-but-one, have 
reached the upper bound \citep{Lin86a}. 

\citet{Lin86a} selected homeostatic constants for  
synapses connecting layers $\layer{A}$ and $\layer{B}$ such that the mean weight 
was unstable and, consequently, all synapses diverged to the upper bound. 
For connections between layers $\layer{B}$ and $\layer{C}$, the homeostatic constants 
are chosen such that the mean weight is stable, requiring some individual synapses 
to diverge to the lower bound and others to the upper bound.  

With synaptic connections between layers $\layer{A}$ and $\layer{B}$ assumed to 
all reach the upper bound, the focus is on determining the learned synaptic 
structure for postsynaptic neurons in layer $\layer{C}$. Given that the learning 
equation in \eq{\ref{eq:learningRule_Linsker_BC}} is linear within the weight 
bounds, the system lends itself to an eigenfunction analysis. That is, we wish 
to identify the independent eigenfunctions that contribute to the evolution of 
synaptic weights. Given that the system is driven by unstructured noise, it 
will self-organise such that the eigenfunction with the leading eigenvalue will 
ultimately dominate the synaptic weight structure. 

In order to conduct an eigenfunction analysis, we approximate the discrete grid 
of neurons by its continuous limit. The probability of a synaptic connection 
existing between neuron $\preN$ at position 
$\icol{ \posx{\preN}{\postN} }{ \posy{\preN}{\postN}}$ in the 
presynaptic layer and postsynaptic neuron $\postN$, detailed in \eq{\ref{eq:2DGauss}}, 
becomes a synaptic density describing the expected proportion of the total number 
of presynaptic inputs originating from 
$\icol{ \posx{\preN}{\postN} }{ \posy{\preN}{\postN}}$. The synaptic strength 
is then considered to be the average weight of synapses at this location. 
In the continuous limit, the learning equation in \eq{\ref{eq:cdot_BCcov}} becomes
\begin{align}\label{eq:learningEqnCont_k2}
  \etalayer{} \Lwgtcont{\poscont}{}
 &= \kone{}{} + 
    \mathop{\mathlarger{\mathlarger{\int}}}_{-\infty}^{\infty} \cxcoeff \; 
       \left( \exp{- \frac{\abs{\poscont -\postwocont}^2}
				                      {2\radiusvar{\inputlayer}{\layer{B}}}} 
		 	+ \ktwo{}{} 
	   \right) 
       \exp{- \frac{\abs{\postwocont}^2 + \abs{\poscont}^2 } 
				   {\radiusvar{\layer{B}}{\layer{C}}}}
       \Lwgtcont{\postwocont}{}
	   d^2 \postwocont \, ,
\end{align}
where neuron $\postN$ in layer $\layer{B}$ is denoted by its continuous position 
vector $\poscont = \icol{\posx{\postN}{\postpostN}}{\posy{\postN}{\postpostN}}$ 
and neuron $\postNtwo$ in layer $\layer{B}$ is represented by its continuous vector, 
$\postwocont = \icol{\posx{\postNtwo}{\postpostN}}{\posy{\postNtwo}{\postpostN}}$, 
where subscripts have been omitted for readability. The Cartesian coordinates have 
been centred on the layer $\layer{C}$ neuron. Note that $\cxcoeff$ contains 
coefficients to normalise covariance and connection probabilities, such that 
$\cxcoeff{} = \left( \pi \radiusvar{\layer{B}}{\layer{C}} \right)^{-2} \, $. 

To characterise the system in terms of its eigenfunctions, we need to solve the 
eigenvalue problem for the system, 
\begin{align}\label{eq:learningEqnCont}
    \eigvalue{}{} \etalayer{} \Lwgtcont{\poscont}{}
 \,& = \, 
    \mathop{\mathlarger{\mathlarger{\int}}}_{-\infty}^{\infty} \cxcoeff \;  
       \left( \exp{- \frac{\abs{\poscont - \postwocont}^2}
				                   {2\radiusvar{\inputlayer}{\layer{B}}}} 
		 	  + \ktwo{}{} 
	   \right) 
       \exp{- \frac{ \abs{\postwocont}^2 + \abs{\poscont}^2}
		           { \radiusvar{\layer{B}}{\layer{C}} }}
       \Lwgtcont{\postwocont}{}
	   d^2 \postwocont \, .
\end{align}

\section{Radial eigenfunctions of the learning equation}\label{sec:radial_eigen}
	Given the circular symmetry of the spatial opponent neurons that emerge from 
	\citepos{Lin86a} network, we derive the radial eigenfunctions and eigenvalues 
	of a layer $\layer{C}$ neuron's receptive field. By identifying the eigenfunction 
	with the largest eigenvalue, we can analytically determine the expected 
	receptive field of the neuron, since this eigenfunction is expected to grow 
	most rapidly and dominate development of the receptive field. 

	\subsection{Radial eigenfunctions of the simplified learning equation}
	   
To proceed we initially set $\ktwo{}{}$ to zero and later consider the more general 
case in which $\ktwo{}{}$ is non-zero. Converting to polar coordinates, such that 
$\posr$ and $\posth$ give the magnitude and phase of $\poscont$, and transforming 
$\posr$ to be unit-less by scaling it by 
$\frac{1}{{\radius{\inputlayer}{\layer{B}}}}$, the eigenvalue problem in 
\eq{\ref{eq:learningEqnCont}} becomes 
\begin{align}\label{eq:eig_k2=0_v1}
    \eigvalue{}{} \etalayer{} \Lwgtcont{\posr}{\posth}
  \, = \, 
 &  \cxcoeff \radiusvar{\inputlayer}{\layer{B}} 
    \exp{- \frac{\posr^2}{2} 
		   \left( \frac{2\radiusvar{\inputlayer}{\layer{B}} + \radiusvar{\layer{B}}{\layer{C}} }
				       { \radiusvar{\layer{B}}{\layer{C}}} 
		   \right)} 
    \mathop{\mathlarger{\mathlarger{\int}}}_{0}^{\infty} d\posintr \posintr 
    \exp{- \frac{\posintr^2} {2}
	  	   \left( \frac{2\radiusvar{\inputlayer}{\layer{B}} + \radiusvar{\layer{B}}{\layer{C}}} 
		               { \radiusvar{\layer{B}}{\layer{C}}} 
		   \right)}  \notag \\
 &  \mathop{\mathlarger{\mathlarger{\int}}}_{0}^{2\pi} d\postwoth
          \exp{- \frac{-2\posr \posintr \cosbr{\posth-\postwoth}{} }
				      { 2}} 
          \Lwgtcont{\posintr}{\postwoth}.
\end{align}

The eigenfunctions and eigenvalues for the simplified learning equation are 
derived in Appendix~\ref{app:radialEigen_simplified}. Introducing a radial decay 
parameter that controls the rate of decay from the centre of the receptive 
field,
\begin{align}\label{eq:eigendecay_radial}
   \eigdecay 
 &= 
   \frac{ \radiusvar{\layer{B}}{\layer{C}} } 
        { 2 \radius{\inputlayer}{\layer{B}} 
			\sqrt{\radiusvar{\inputlayer}{\layer{B}} + \radiusvar{\layer{B}}{\layer{C}} }},
\end{align}
the eigenfunctions and associated eigenvalues can be expressed in polar 
coordinates as
\begin{subequations}\label{eq:simpleEigenSoln}
\begin{align}
    \eigvalue{\fouriersum}{\eigLindex} 
 &= 
    2 \pi \cxcoeff 
    \left( \frac{\eigdecay   \radiusvar{\layer{B}}{\layer{C}} }
			    {\eigdecay ( \radiusvar{\inputlayer}{\layer{B}} 
						  + \radiusvar{\layer{B}}{\layer{C}} )
			   + \radiusvar{\layer{B}}{\layer{C}} }
	\right)^{\fouriersum+\eigLindex+1}			        \label{eq:simpleEigenvalue} \\
    \eigvec{\fouriersum,\eigLindex}{\posr,\posth}
 &= 
    \eignorm{\fouriersum}{\eigLindex} \posr ^{\fouriersum-\eigLindex} 
    \exp{-\frac{\posr^2}{2\eigdecay}} 
    \laguerre{\eigLindex}{\fouriersum-\eigLindex}{\frac{\posr^2}{\eigdecay}}
	\exp{ \imag (\fouriersum-\eigLindex) \posth } \, ,  \label{eq:simpleEigenfn} 
\end{align}
\end{subequations}
where $\eignorm{\fouriersum}{\eigLindex}$ is a normalisation factor and 
$\laguerre{\eigLindex}{\fouriersum-\eigLindex}{}$ is an associated Laguerre polynomial. 
Since $\int_0^\infty x^p e^{-x}\mathrm{L}_q^p(x)^2\,dx=(p+q)!/q!$, the normalisation 
factor can be derived as
\begin{align}
    \eignorm{\fouriersum}{\eigLindex} 
 &= 
    \begin{cases}
		\sqrt{ \frac{2}{\pi \eigdecay
					    \radiusvar{\inputlayer}{\layer{B}} } }, 
		\qquad &\fouriersum=\eigLindex \\
		\sqrt{ \frac{\eigLindex!}
					{\pi \fouriersum! \eigdecay^{\fouriersum-\eigLindex+1}
					 \radiusvar{\inputlayer}{\layer{B}} } },
		\qquad &\textrm{otherwise},
	\end{cases}
\end{align}
where the factor of $2$ difference occurs for the case $\fouriersum=\eigLindex$ 
because the integral for the angular component is over $\cos{(0\posth)}$, a constant. 

\begin{figure}[ht!]
\centering
	\includegraphics[width=0.8\textwidth]{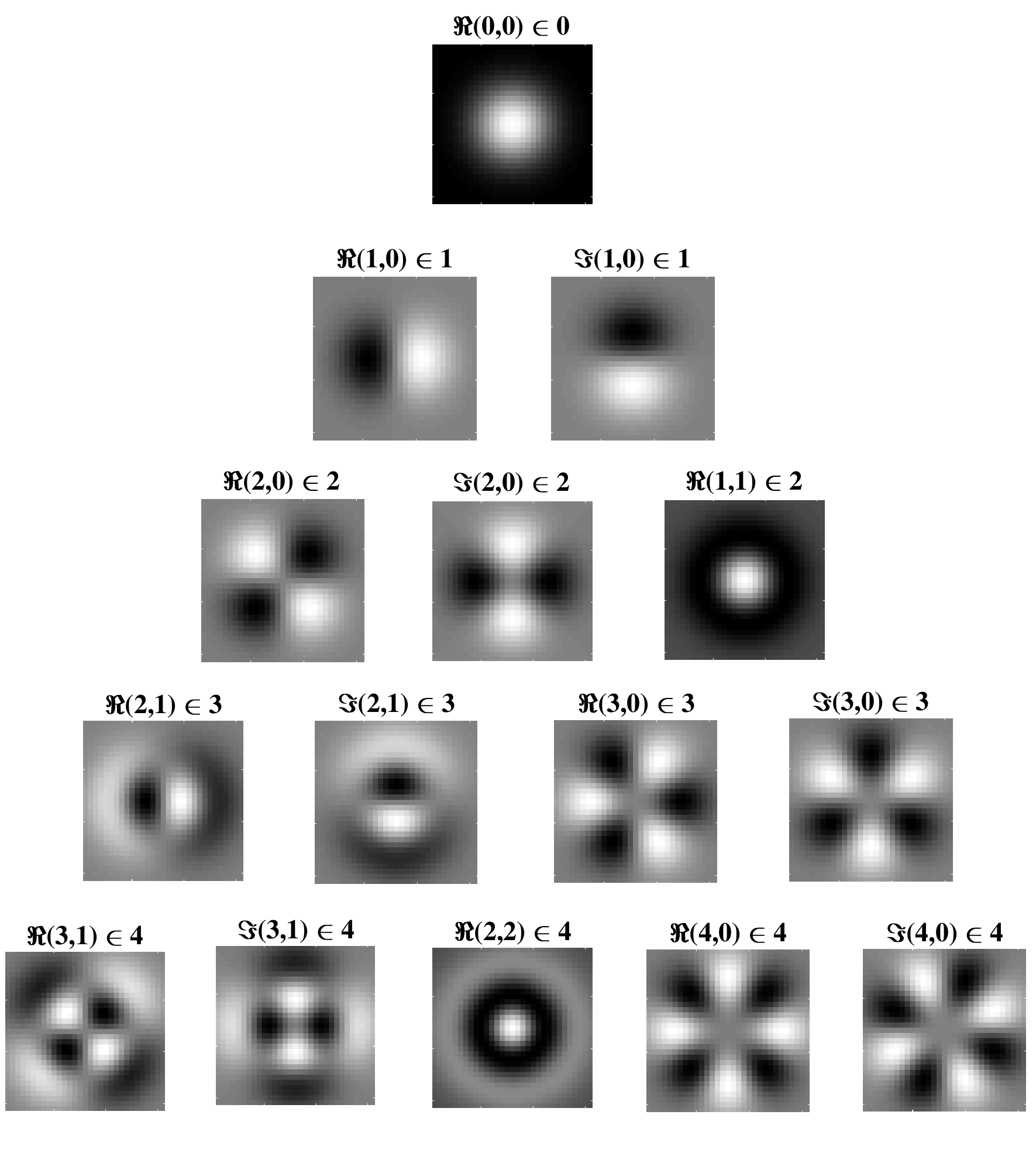}
	\caption{Eigenvalues and eigenfunctions of the simplified learning equation,  
		     \eq{\ref{eq:eig_k2=0_v1}}, given in \eq{\ref{eq:simpleEigenSoln}} 
			 for pairs of indices, 
			 $\icol{\fouriersum}{\eigLindex} \in \lambda_{\fouriersum,\eigLindex}$. 
			 Eigenvalues are ordered by $\fouriersum + \eigLindex$, with 
			 $\fouriersum + \eigLindex = 0$ giving the largest eigenvalue and, 
			 hence, $\icol{0}{0}$ being the leading eigenfunction. Eigenfunctions 
			 in the same row have the same eigenvalue and are, therefore, degenerate. 
			 Eigenfunctions are given for both the real part of 
			 \eq{\ref{eq:simpleEigenfn}} (i.e., the cosine angular component) and 
			 for the imaginary part, denoted by $\mathfrak{R}$ and $\mathfrak{I}$, 
			 respectively. From the figure, it can be seen that, when 
			 $\fouriersum=\eigLindex$, the eigenfunction is radially symmetric, 
			 being fully determined by the radial component of the eigenfunction. 
			 White indicates positive regions of synaptic weights, while black 
			 indicates negative regions. The leading eigenfunction is all positive. 
			 As $\fouriersum-\eigLindex$ increases, the frequency 
			 of the angular component increases. 
		     \label{fig:radialEigenfunctions}}
\end{figure}

Eigenfunctions up to order $4$ are shown in \fig{\ref{fig:radialEigenfunctions}} 
in order of decreasing eigenvalue, $\eigvalue{}{}$. The eigenfunctions are ordered 
by $\fouriersum + \eigLindex$, where $\eigLindex$ controls the shape of the 
Laguerre polynomial and $\fouriersum-\eigLindex$ controls the angular frequency. 
The eigenfunction with the 
largest eigenvalue has order $\fouriersum + \eigLindex = 0$ and is radially 
symmetric with all positive synaptic weights. Consequently, for the simplified 
learning equation described in \eq{\ref{eq:eig_k2=0_v1}} and after learning for a 
sufficiently long period, the synaptic weight structure of a layer $\layer{C}$ 
postsynaptic neuron will be all excitatory connections with weights at the upper 
bound. 

For completeness, we derive the eigenfunctions and eigenvalues of \citepos{Lin86a} 
network using Cartesian coordinates, the solution of which is a special case of 
that found in \citet{WimWenMilvanH97b}. We show that a weighted sum of the 
Cartesian eigenfunctions produces the radial eigenfunctions, thus establishing 
equivalence. The derivations are given in Appendix~{\ref{app:CartesianEigen}}. 
For eigenvalues indexed by order $\eigxindex$ and $\eigyindex$, for the $x$ 
and $y$ dimensions respectively, eigenfunction and eigenvalue pairs are given by 
\begin{subequations}\label{eq:eig_k2=0}
\begin{align}
   \eigvalue{\eigxindex}{\eigyindex}
 &= 
   2\pi\radiusvar{\inputlayer}{\layer{B}} q^{\eigxindex + \eigyindex + 1} 
   \label{eq:cartesianEigenval_simple} \\
    \eigvec{\eigxindex,\eigyindex}
         {\frac{\posxcont}{\sqrt{\eigdecay}}, \frac{\posycont}{\sqrt{\eigdecay}}}
 &= 
    \hermitecoeff{\eigxindex} \hermitecoeff{\eigyindex} 
    \hermite{\eigxindex}{\frac{\posxcont}{\sqrt{\eigdecay}}} 
    \hermite{\eigyindex}{\frac{\posycont}{\sqrt{\eigdecay}}} 
	\exp{-\frac{\posxcont^2 + \posycont^2}{2\eigdecay}} \, . 
   \label{eq:cartesianEigenfn_simple} 
\end{align}
\end{subequations}

\begin{figure}[ht!]
\centering
	\includegraphics[width=0.8\textwidth]{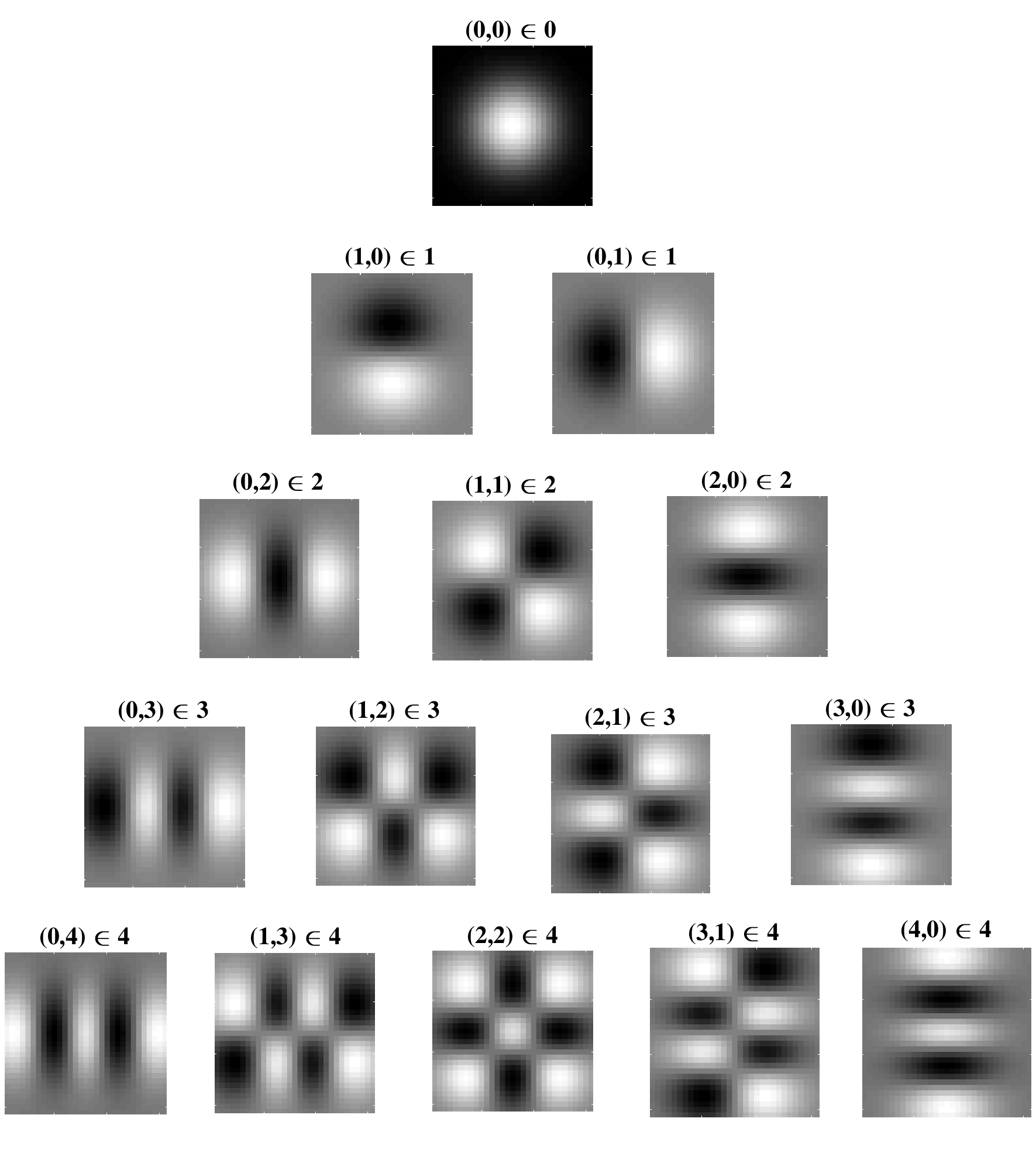}
	\caption{Cartesian eigenfunctions of the simplified learning equation,  
		     \eq{\ref{eq:Linsker_eigfnLearningEqn}}, defined for index pairs 
			 $\left(\eigxindex, \eigyindex\right)$. Eigenvalues are 
			 determined by $\eigxindex + \eigyindex$, where $\left(0,0\right)$ 
			 has the largest eigenvalue and therefore 
			 $\eigvec{0,0}{\frac{\posxcont}{\sqrt{\eigdecay}}, 
			               \frac{\posycont}{\sqrt{\eigdecay}}}$
			 is the leading eigenfunction. Eigenfunctions in the same row have 
			 the same eigenvalue and are, therefore, degenerate. Eigenvalues 
			 decrease with descending rows. Regions of white indicate 
			 positive synaptic weights, while black indicates negative weights. The leading 
			 eigenfunction has all positive synapse weights. 
			 \label{fig:CartesianEigenvec}}
\end{figure}

\fig{\ref{fig:CartesianEigenvec}} shows Cartesian eigenfunctions up to the fourth 
order, which is determined by $\eigxindex + \eigyindex$. The eigenfunctions are 
shown in order of decreasing eigenvalue, so that the eigenfunction with the largest 
eigenvalue is of order $\eigxindex + \eigyindex = 0$. This eigenfunction is radially 
symmetric, with all positive weights. Consequently, the Cartesian eigenfunctions 
of the simplified learning equation described in \eq{\ref{eq:Linsker_eigfnLearningEqn}} 
give the same result as the radial eigenfunctions, shown in 
\fig{\ref{fig:radialEigenfunctions}}. After sustained learning, the weight structure 
of a neuron in layer $\layer{C}$ will have all synapses at the upper bound. 

\begin{figure}[ht!]
\centering
	\includegraphics[width=0.8\textwidth]{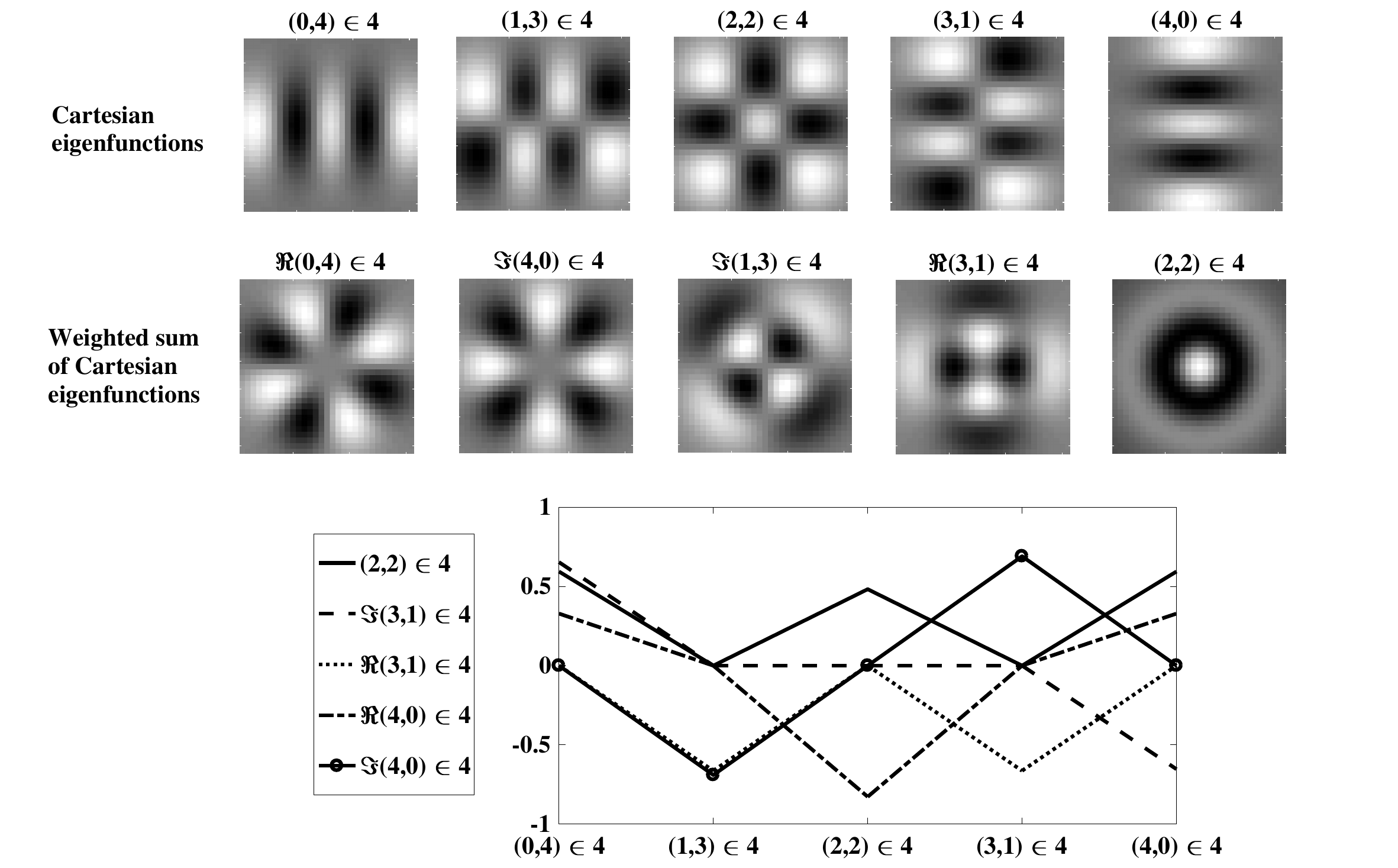}
	\caption{A weighted sum of degenerate Cartesian eigenfunctions of order 
		$\eigxindex + \eigyindex = 4$ are used to generate radial eigenfunctions 
		of order $\fouriersum + \eigLindex = 4$. Weights are determined using 
		maximum likelihood regression. 
		The top row shows degenerate Cartesian eigenfunctions of order $4$. 
		The middle row shows the weighted sum of the Cartesian eigenfunctions, 
		reproducing the radial eigenfunctions in \fig{\ref{fig:radialEigenfunctions}}. 
		The bottom row shows the regression weights. 
		\label{fig:CartesianRegression}}
\end{figure}


	\subsection{Radial eigenfunctions of the full learning equation}
	   
While covariance between the activity of layer $\layer{B}$ input neurons primarily 
drives the structure of the layer $\layer{C}$ cell, the 
$\kone{\layer{B}}{\layer{C}}$ and $\ktwo{\layer{B}}{\layer{C}}$
terms control the homeostatic equilibrium. \citet{MacMil90} empirically showed that 
the choice of $\ktwo{\layer{B}}{\layer{C}}$ can change the structure of the dominant 
eigenfunction, and hence the resultant receptive field of a layer $\layer{C}$ cell. 
As \fig{\ref{fig:radialEigenfunctions}} shows, for the simplified system, the leading 
eigenvalue has all synapses at the upper or the lower bound. For a negative value 
of $\ktwo{\layer{B}}{\layer{C}}$, homeostasis can only be reached if some of the 
synapses are negative. To determine the impact of the learning constant, we find 
an analytical expression for the eigenfunctions of the full learning equation, 
\eq{\ref{eq:learningEqnCont}}, by conducting a perturbation analysis on the 
simplified learning equation, in \eq{\ref{eq:learningEqnCont}}. 

The full derivation is detailed in Appendix~\ref{app:radialEigen_full}. The 
eigenfunctions of the first order perturbation are equal to those of the 
simplified equation, \eq{\ref{eq:simpleEigenfn}}. However, the eigenvalues 
are altered by the addition of the learning constants according to 
\begin{align}\label{eq:lambda_k2}
    \eigvaluePerturb{\fouriersum}{\eigLindex} 
  &=
    \eigvalue{\fouriersum}{\eigLindex} + W_{\fouriersum,\eigLindex}, 
\end{align}
where 
\begin{align}\label{eq:eigvalue_perturb}
   W_{\fouriersum,\eigLindex}
 &= 
   \pi \eigdecay^{\fouriersum-\eigLindex+1} \ktwo{\layer{B}}{\layer{C}} 
   \eignorm{\fouriersum}{\eigLindex}^2
   \frac{ \Gamma\left(\fouriersum + \eigLindex + 1\right)
          \left( \alpha -1 \right)^{2\eigLindex} }
		{ \eigLindex!^2 \alpha^{\fouriersum + \eigLindex + 1} }
    { }_2F_1 \left( -\eigLindex, \eigLindex;\: \fouriersum - \eigLindex;\: 
		              \frac{ \alpha (\alpha-2) }{ (\alpha-1)^2 } 
		     \right) ,
\end{align}
and ${}_2F_1\left( \right)$ is the hypergeometric function. 
As detailed in Appendix~\ref{app:radialEigen_full}, the only non-zero perturbations 
are where $\fouriersum+\eigLindex$ is even and $\fouriersum=\eigLindex$, which 
happens only once for each even order degenerate eigenfunction set. 

Inspection of \eq{\ref{eq:eigvalue_perturb}} reveals that, for positive $\ktwo{}{}$, 
perturbation of the eigenvalues is positive and monotonically decreasing with 
$\fouriersum + \eigLindex$. Consequently, the order of the eigenvalues remains 
the same. For negative $\ktwo{}{}$, the perturbation on the eigenvalues is 
negative and monotonically increasing with eigenfunction order, 
$\fouriersum + \eigLindex$. Since these perturbations are being added 
to the original eigenvalues, which are positive, the result can be a change in 
the dominant eigenfunction. This result supports the empirical findings by 
\citet{MacMil90} who showed the emergence of a spatial opponent cell in 
$\layer{C}$, where $\fouriersum+\eigLindex = 0$ for small values of $\ktwo{}{}$, 
and bi-lobed neurons with $\fouriersum + \eigLindex = 1$, for larger values of 
$\ktwo{}{}$. 

\begin{figure}[ht!]
\centering
	\includegraphics[width=0.6\textwidth]{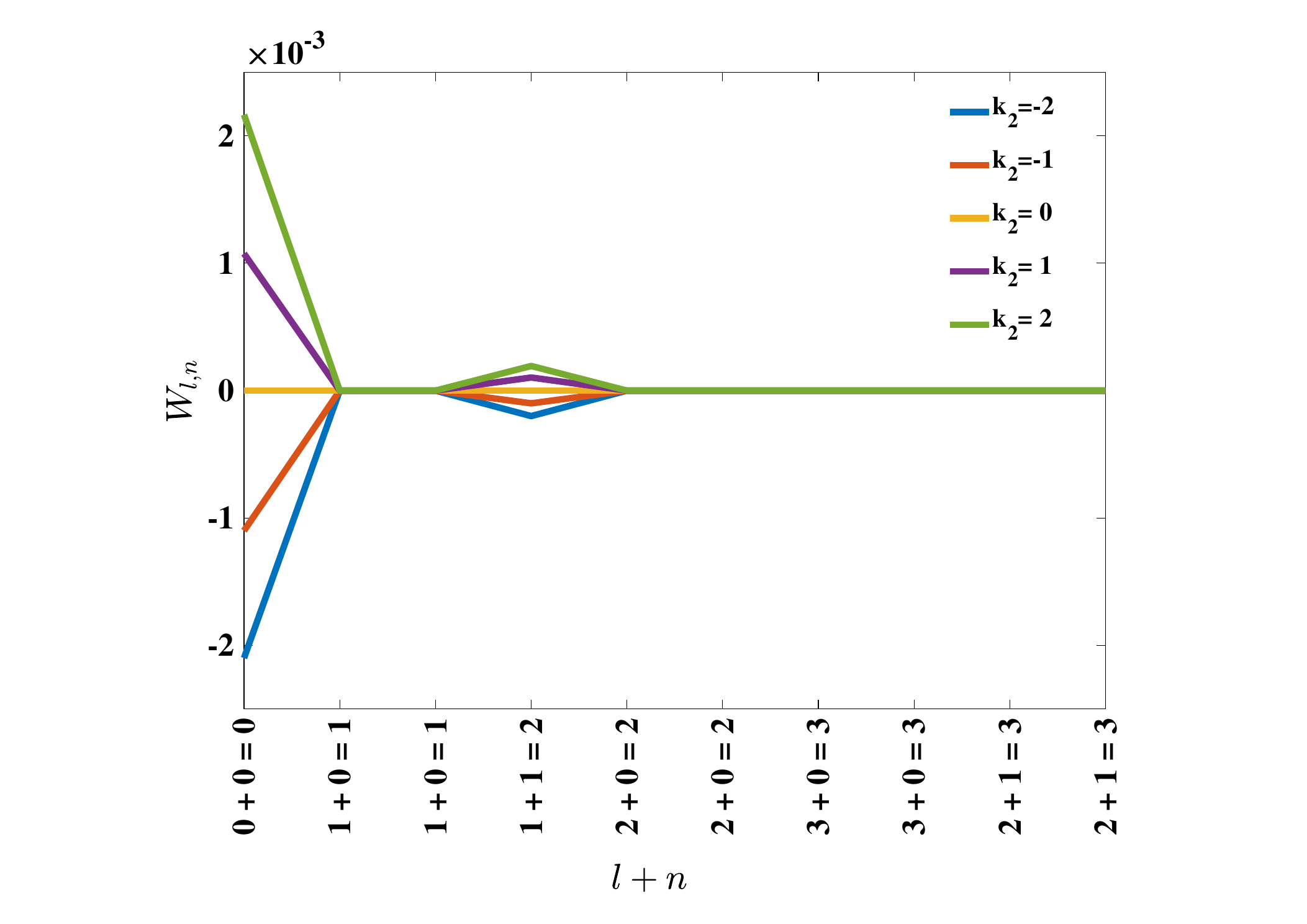}
	\caption{Effect of adding the perturbation term, $\ktwo{}{}$, on eigenvalues 
			 $\lambda_{\fouriersum,\eigLindex}$, represented by 
			 $W_{\fouriersum,\eigLindex}$ in \eq{\ref{eq:lambda_k2}}. For positive 
			 $\ktwo{}{}$, the perturbation results in $W_{\fouriersum,\eigLindex}$ 
			 being positive, while for negative $\ktwo{}{}$, the perturbation 
			 causes $W_{\fouriersum,\eigLindex}$ to be negative. 
		    \label{fig:perturb_k2}}
\end{figure}

	\section{Emergence of radial orientation selectivity}\label{sec:emergence_radial}
		The original network proposed by \citet{Lin86a}, and for which we have 
calculated the eigenfunctions, made an implicit assumption that neurons within 
each layer were evenly distributed and that receptive fields of all neurons 
in a layer were statistically identical, drawn from the same synaptic 
connectivity distribution. However, it is known that some biological neuron 
layers show an uneven density of cells across the lamina, and contain 
receptive fields with different statistical properties. 

We consider the impact of changing neuron density as a function of 
distance to the layer centre. We assume that a consequence of this is that 
the radius of a neuron's synaptic connectivity distribution becomes dependent 
on the neuron's position in the layer. That is, where neurons are spread further 
apart there is an increase in connectivity radius to compensate. 
For simplicity assume that a postsynaptic 
neurons's arbor is within a sufficiently small area that the presynaptic neuron 
connection density is parameterised by a constant radius. If we denote the 
spatial centre of the  neuron layer by $\centreN{}$ and consider this point to 
have location vector $\posvec{0}{0}$, then a postsynaptic cell, located at 
$\posvec{\posx{\postN}{\centreN}}{\posy{\postN}{\centreN}}$, 
has connection density that is a function of the magnitude of its position, 
\begin{align}\label{eq:radial_distance}
   \distance{\postN}{\centreN}{\layer{B}}{\layer{B}}
 &=
   \left( \posx{}{\postN}^2 + \posy{}{\postN}^2 \right)^{\nicefrac{1}{2}}. 
\end{align}

Let the radius of a cell be a linear function of its radial distance to the layer 
centre, such that
\begin{align}
    \radius{\layer{A}}{\layer{B}}_{\postN} 
 &= \distance{\postN}{\centreN}{\layer{B}}{\layer{B}} 
					 \radius{\layer{A}}{\layer{B}}. 
\end{align}

In this scenario, the probability of presynaptic neuron, $\preN$, in layer 
\layer{\inputlayer}, generating a synaptic connection to postsynaptic neuron, 
$\postN$, in layer $\layer{B}$, is given by 
\begin{align}\label{eq:2DGauss_radial}
	\probdist{N}{ \icol{ \posx{\preN}{\postN} }{ \posy{\preN}{\postN}} }
				{ \frac{1}{2} \radiusvar{\layer{A}}{\layer{B}} }
  &=
    \frac{1}{\pi (\distance{\postN}{\centreN}{\layer{B}}{\layer{B}})^2 
		     \radiusvar{\layer{A}}{\layer{B}} }  
	\exp{- \frac{ \posx{\preN}{\postN}^2 + \posy{\preN}{\postN}^2}
			    {(\distance{\postN}{\centreN}{\layer{B}}{\layer{B}})^2 
				  \radiusvar{\layer{A}}{\layer{B}} } }.
\end{align}

In Appendix~\ref{app:sharedInput_radial}, we calculate the expected number of shared 
inputs between two neurons in layer $\layer{B}$. This is important to consider as 
shared inputs is the source of correlation between layer $\layer{B}$ neurons, which 
then triggers the emergence of spatial opponent neurons in \citepos{Lin86a} network. 
In the case of the synaptic connection radius increasing linearly with distance 
from the centre of the neuron layer, the expected number of shared inputs between 
two layer $\layer{B}$ neurons is found to be
\begin{align}\label{eq:radialSharedConns}
   \expect{\Nshared{\layer{A}}{\layer{B}}; 
	       \posvec{\posx{\postN}{}}{\posy{\postN}{}},
	       \posvec{\posx{\postNtwo}{}}{\posy{\postNtwo}{}} }
  =&
   \frac{(\N{\layer{A}}{\layer{B}})^2}
		{\pi \radiusvar{\layer{A}}{\layer{B}}
         \left(\distance{\postN}{}{}{}^2 + \distance{\postNtwo}{}{}{}^2 \right)}
   \exp{-\frac{\distance{\postN}{\postNtwo}{}{}^2}
			  {\radiusvar{\layer{A}}{\layer{B}}
			   \left(\distance{\postN}{}{}{}^2 
			       + \distance{\postNtwo}{}{}{}^2\right)}}.
\end{align}

We simulated this learning equation and plotted the receptive fields of three 
layer $\layer{C}$ neurons in three different positions relative to a small 
layer of $\layer{B}$ neurons. 
The neurons developed radial orientation tuning, with tuning curves showing 
an orientation directed towards the centre of the lamina. 

\begin{figure}[ht!]
\centering
	\includegraphics[width=0.9\textwidth]{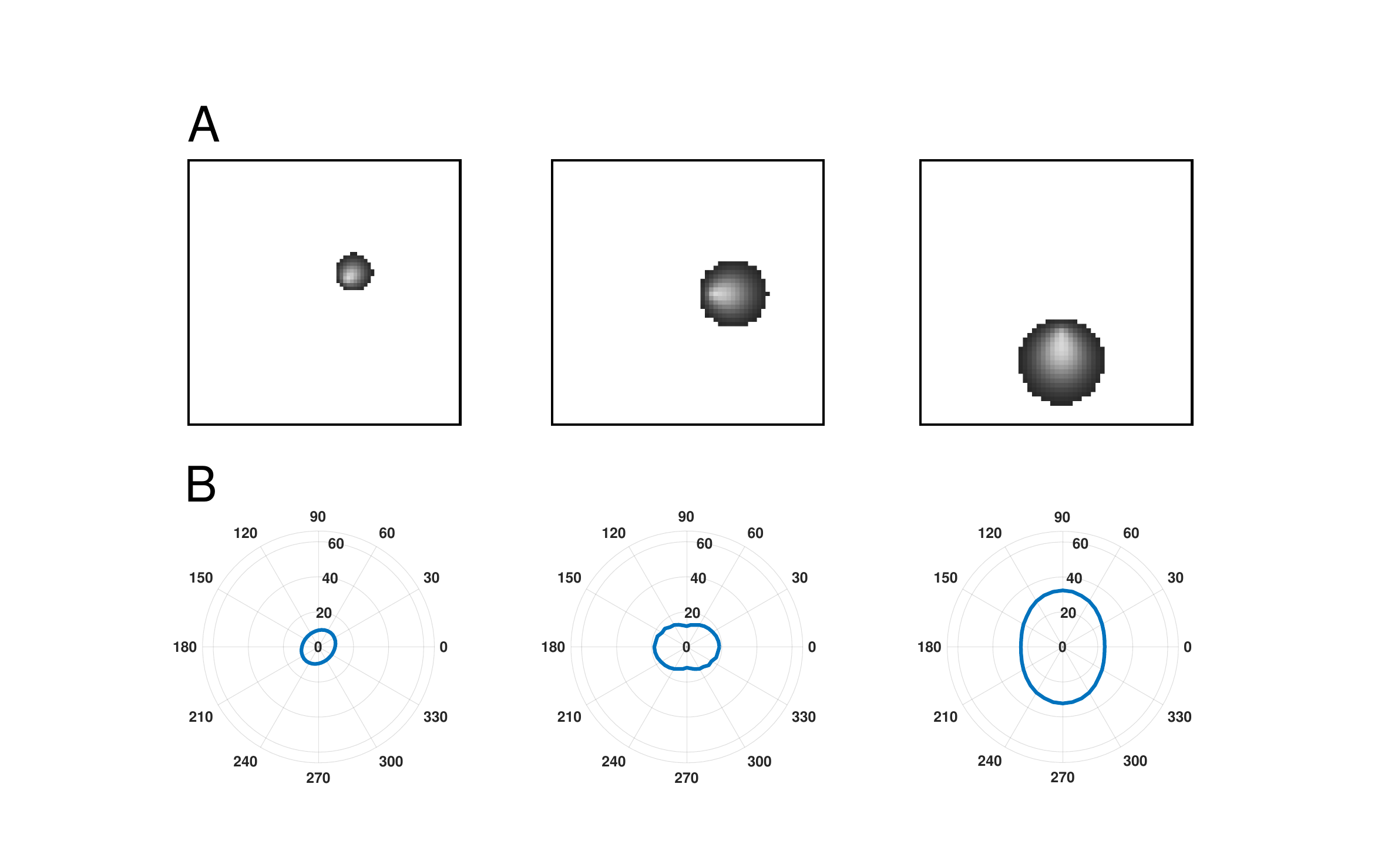}
	\caption{Examples of receptive fields and tuning curves.
	    \figpanel{A} 
		Receptive fields predicted by the leading eigenfunction of three 
		layer $\layer{C}$ neurons at different locations relative to the centre 
		of layer $\layer{B}$. The layer is small relative to the receptive field 
		size of the layer $\layer{C}$ neurons to highlight changes in receptive 
		field size and the radial orientation selectivity of selected layer 
		$\layer{C}$ neurons. These features emerge from a linearly increasing arbor 
		radius in synaptic connectivity between layer $\layer{A}$ and layer 
		$\layer{B}$ neurons. As established by \citet{Lin86a}, it is the overlap 
		between the arbors of layer $\layer{B}$ neurons that prompts correlation 
		in their activity despite only inputting unstructured noise into layer 
		$\layer{A}$ and, subsequently, generates the structured receptive 
		fields found in the layer $\layer{C}$ neurons. 
		\figpanel{B} 
		Tuning curves generated for each of the layer $\layer{C}$ neurons in 
		panel A. Tuning curves were calculated for the spatial frequency 
		prompting the largest response in the neuron. 
		\label{fig:radialTuning}}
\end{figure}

	\FloatBarrier

\section{Discussion}\label{sec:discussion}
	   
In this paper, we provide a general expression for the complete set of eigenfunctions 
for the three-layer feed-forward network proposed by \cite{Lin86a}. Initially, 
the homeostatic parameters were set to zero to simplify the learning equation. 
This result was then extended via a perturbation analysis to provide the complete 
set of eigenfunctions for the network with non-zero homeostatic parameters. 

\citepos{Lin86a} analysis was integral in revealing how neural learning occurred 
before the onset of structured environmental input, empirically demonstrating
the emergence of spatial opponent neurons in the third layer. \citet{MacMil90} 
provided a stability analysis of \citepos{Lin86a} network, noting the first 
six eigenfunctions, determined via an ansatz based on the results of 
\citet{Tang90}. \citet{MacMil90} showed that the receptive field structure of 
cells in the third layer could be either spatial opponent neuron or bi-lobed neurons, 
depending upon the value of the homeostatic parameters. Similarly, \citet{WalBis92} 
extended \citepos{Lin86a} network to the auditory system, considering the morphology 
of the resulting neuron based on the homeostatic parameters of the system. In this 
paper, we provide the complete set of eigenvalues for the full learning equation, 
enabling an exact calculation of the homeostatic parameters required to induce 
this change, and a quantitative analysis on the parameter space. 

\cite{Lin86b} showed that augmenting the network with additional layers prompts 
the development of orientation selective neurons. However, given the absence of a
complete mathematical framework for the three-layer network, there has been
limited mathematical analysis provided for the development of orientation selective 
neurons in \citepos{Lin86b} network. \citet{Yam02} provided an analysis of deeper
layers, essentially based on an ansatz for the eigenfunctions for the three-layer
network. The results in this paper provide the foundation for analysis of larger
networks and hence the development of features other than spatial opponent neurons. 
As the system is radially symmetric in connectivity distribution, radial
eigenfunctions provide a natural coordinate system that will facilitate future
work on more complex network and parameter regimes.

The results of this study demonstrate that relaxing the assumption of evenly 
distributed neurons across the layer can change the receptive fields that emerge 
in the third layer. Similar to the distribution in the retina, we examined a 
decrease in neuron density with increasing distance from the centre of the layer 
and, consequently, an increase in synaptic connectivity radius. We analytically 
derived an expression for the learning equation in the third layer, as a result 
of a radially dependent connectivity distribution between the first and second 
layers.  The eigenfunctions for the learning equation were empirically calculated, 
showing that orientation selective neurons emerge. Interestingly, the preferred 
orientation of the neurons was the radial orientation towards the centre of the 
laminar. 

It is well established that neural density changes as a function of eccentricity 
in the retina \citep{SjoOlsPopCon99,Wat14}, and receptive field sizes of neurons 
in the primary visual cortex increase with stimulus eccentricity 
\citep{WurMinYaz13,SmiSinWilGre01}. Furthermore, it is known that orientation 
selectivity in the primary visual cortex is biased towards radial orientation, 
in that an orientation selective neuron in the primary visual cortex is more 
likely to be oriented towards the centre of the retina \citep{RodRevPig04}. Our 
results provide a mechanism, based upon the network configuration and neural 
plasticity, to account for this observed emergence of radially oriented cells.

\section*{Acknowledgements}

	The authors acknowledge support under the Australian Research Council (ARC) Discovery 
	Projects funding scheme (Project DP140102947).
	Carlo Beenakker is acknowledged for assistance in evaluating the integral in 
	\eq{\ref{eq:general_integral}}. 

\appendix
\section*{Appendix}

   \section{Expected number of shared inputs}\label{app:sharedInputs}
	   
To examine network dynamics, it is necessary to ascertain the expected number of 
shared connections between two neurons. The number of shared connections from a 
presynaptic layer to two neurons in the postsynaptic layer, say $\postN$ and 
$\postNtwo$, depends on the radial distance between them since the synaptic 
connection density for each is a Gaussian function of distance 
(see \fig{\ref{fig:network}}). We assume for simplicity and without loss of 
generality that $\postN$ and $\postNtwo$ differ only in their $\posx{}{}$ 
coordinate so that 
$ \distance{\postN}{\postNtwo}{\layer{B}}{\layer{B}} 
= \posx{\preN}{\postN} - \posx{\preN}{\postNtwo}$.

Center the Cartesian coordinates describing a neuron's position in the 
laminar on one of the postsynaptic neurons, say $\postN$, so that the other 
postsynaptic neuron, say $\postNtwo$, lies on the $x$ axis. From 
\eq{\ref{eq:2DGauss}}, neuron, $\preN$, 
in layer $\inputlayer$, has a probability of connecting to neuron $\postN$ 
in layer \layer{B} of 
$\probdist{N}{\posx{\preN}{\postN},\posy{\preN}{\postN}}
             {\vect{0},\Sigma^{\inputlayer}}$ 
and a probability of connecting to neuron $\postNtwo$ in layer $\layer{B}$ of 
$ \probdist{N}{\posx{\preN}{\postNtwo},\posy{\preN}{\postNtwo}}
              {\vect{0},\Sigma^{\inputlayer}}
= \probdist{N}{\posx{\preN}{\postN} - 
	           \distance{\postN}{\postNtwo}{\layer{B}}{\layer{B}},
	           \posy{\preN}{\postN}}
              {\vect{0},\Sigma^{\inputlayer}}$.
The probability of the presynaptic neuron connecting to both postsynaptic neurons 
$\postN$ and $\postNtwo$ is simply the product of the probability of each individual 
connection being made. The expected number of common connections can be determined 
by summing this joint probability over the layer or, in the continuous limit, 
integrating the joint probability over the layer of presynaptic 
neurons. If $\N{\inputlayer}{\layer{B}}$ denotes the number of synaptic connections 
from layer $\inputlayer$ to a layer $\layer{B}$ neuron and 
$\atDist{\Nshared{\layer{B}}{}}$ the number of shared connections between 
two postsynaptic neurons in layer $\layer{B}$ separated by a distance of 
$\distance{}{}{}{}$, then in the continuous limit, 
\begin{align}
    \atDist{\Nshared{\layer{B}}{}}
 &= (\N{\inputlayer}{\layer{B}})^2   
    {\mathlarger{\iint\limits_{\posx{}{}\posy{}{}}}}
    \probdist{N}{\posx{}{},\posy{}{}}
                {\vect{0},\Sigma^{\inputlayer}}
    \probdist{N}{\posx{}{} - \distance{}{}{}{},\posy{}{}}
                {\vect{0},\Sigma^{\inputlayer}}
    \,d\posx{}{} \,d\posy{}{}, 
\end{align}
where the sub- and super-scripts on distance parameters have been dropped to aid 
readability. This can be expanded as  
\begin{align}
   \atDist{\Nshared{\layer{B}}{}}
 &= 
  (\N{\inputlayer}{\layer{B}})^2
  {\mathlarger{\iint\limits_{\posx{}{}\posy{}{}}}}
	 \frac{1}{\left(\pi\radiusvar{\inputlayer}{\layer{B}}\right)^2} 
	 \exp{- \frac{\posx{}{}^2 + \posy{}{}^2}
			           {\radiusvar{\inputlayer}{\layer{B}}} }
	 \exp{- \frac{\left(\posx{}{}-\distance{}{}{}{}\right)^2 + \posy{}{}^2} 
			     {\radiusvar{\inputlayer}{\layer{B}}} }
   \,d\posx{}{} \,d\posy{}{}    \notag \\
 &= 
  \frac{(\N{\inputlayer}{\layer{B}})^2}{\left(\pi\radiusvar{\inputlayer}{\layer{B}}\right)^2} 
  {\mathlarger{\iint\limits_{\posx{}{}\posy{}{}}}}
   \exp{- \frac{ 2 \posx{}{}^2 + 2\posy{}{}^2 
				 + \distance{}{}{}{}^2
				 - 2\posx{}{}\distance{}{}{}{} }
			   {\radiusvar{\inputlayer}{\layer{B}}} }
   \,d\posx{}{} \,d\posy{}{}  \notag   \\
 &= 
  \frac{(\N{\inputlayer}{\layer{B}})^2}{\left(\pi\radiusvar{\inputlayer}{\layer{B}}\right)^2} 
  {\mathlarger{\iint\limits_{\posx{}{}\posy{}{}}}}
	 \exp{-\frac{ 2 \left( \left(\posx{}{}-\frac{\distance{}{}{}{}}{2}\right)^2 
				               + \posy{}{}^2 + \frac{\distance{}{}{}{}^2}{4}
						  \right)}
			           {\radiusvar{\inputlayer}{\layer{B}}} }
   \,d\posx{}{} \,d\posy{}{}.
\end{align}
Introduce $\posx{}{}' = \posx{}{} - \nicefrac{\distance{}{}{}{}}{2}$, so that 
\begin{align}\label{eq:sharedConns}
   \atDist{\Nshared{\layer{B}}{}}
 &= 
   \exp{-\frac{\distance{}{}{}{}^2}{2\radiusvar{\inputlayer}{\layer{B}}}}
   \frac{(\N{\inputlayer}{\layer{B}})^2}{\left(\pi\radiusvar{\inputlayer}{\layer{B}}\right)^2} 
  {\mathlarger{\iint\limits_{\posx{}{}\posy{}{}}}}
   \exp{-\frac{ 2 \left( \posx{}{}'^2 + \posy{}{}^2 \right)}
			  {\radiusvar{\inputlayer}{\layer{B}}} }
   \,d\posx{}{} \,d\posy{}{}                                         \notag \\
 &= 
   \exp{-\frac{\distance{}{}{}{}^2}{2\radiusvar{\inputlayer}{\layer{B}}}}
   \frac{(\N{\inputlayer}{\layer{B}})^2}{\left(\pi\radiusvar{\inputlayer}{\layer{B}}\right)^2} 
   \sqrt{\frac{\pi\radiusvar{\inputlayer}{\layer{B}}}{2}}
   \sqrt{\frac{\pi\radiusvar{\inputlayer}{\layer{B}}}{2}}	                     \notag \\
 &=
   \frac{(\N{\inputlayer}{\layer{B}})^2}{2\pi\radiusvar{\inputlayer}{\layer{B}}}
   \exp{-\frac{\distance{}{}{}{}^2}{2\radiusvar{\inputlayer}{\layer{B}}}},
\end{align}
using the identity
$\int_{-\infty}^{\infty}\exp{-ax^2}=\sqrt{\nicefrac{\pi}{a}}$.

This result demonstrates that the number of shared connections between two neurons 
with Gaussian synaptic connection densities is itself a Gaussian function of the 
radial distance between the neurons with a variance that is half the value of 
the synaptic connection density radius. This means that a postsynaptic neuron 
is expected to have the most common connections with itself, for which 
$\distance{}{}{}{}=0$. Additionally, for small variance or connection radius, 
a postsynaptic neuron will share many connections with proximate neighbors, with the 
number of shared connections falling off quickly with distance. Since the expected 
number of synaptic inputs is constant, a large connection radius implies that the 
neuron will have shared connections with neurons comparatively distal to it, since 
nearby neurons will have comparatively fewer shared connections.

   \section{Covariance of neural activity in layer $\layer{B}$}\label{app:covLayerB}
	   
Expressions for the covariance of layer $\layer{B}$ neurons are derived here. 
Sample covariance between two postsynaptic neuron rates in layer $\layer{B}$, say 
$\Lrate{\postN}{\layer{B}}{}$ and $\Lrate{\postNtwo}{\layer{B}}{}$, for neurons 
$\postN$ and $\postNtwo$, respectively, is calculated as 
\begin{align}
   \scov{\Lrate{\postN}{\layer{B}}{}}{\Lrate{\postNtwo}{\layer{B}}{}}
 &= 
   \expect{\Lrate{\postN}{\layer{B}}{} \Lrate{\postNtwo}{\layer{B}}{}}
 - \expect{\Lrate{\postN}{\layer{B}}{}} \expect{\Lrate{\postNtwo}{\layer{B}}{}}.
 \end{align}
For unitary weights from layer $\layer{A}$ to layer $\layer{B}$,  
\eq{\ref{eq:layerB_rate}}, can be employed to give 
\begin{align}
   \scov{ \Lrate{\postN}{\layer{B}}{} }{ \Lrate{\postNtwo}{\layer{B}}{} }
 &= 
   \expect{ \left(\Ra{\layer{B}} + \Rb{\layer{A}}{\layer{B}}
		       \sum\limits_{\preN} \Lrate{\preN}{\layer{A}}{} 
		    \right)
            \left( \Ra{\layer{B}} + \Rb{\layer{A}}{\layer{B}}
			   \sum\limits_{\preNtwo} \Lrate{\preNtwo}{\layer{A}}{}
			\right) }
 - \expect{\Ra{\layer{B}} + \Rb{\layer{A}}{\layer{B}}
   \sum\limits_{\preN} \Lrate{\preN}{\layer{A}}{}} 
   \expect{\Ra{\layer{B}} + \Rb{\layer{A}}{\layer{B}}
   \sum\limits_{\preNtwo} \Lrate{\preNtwo}{\layer{A}}{}} \notag \\
 &= 
   (\Ra{\layer{B}})^2 + 2\Ra{\layer{B}} \Rb{\layer{A}}{\layer{B}} \meanrate{\layer{B}} 
 + \RbSq{\layer{A}}{\layer{B}}
   \expect{\sum\limits_{\preN}\sum\limits_{\preNtwo}
	           \Lrate{\preN}{\layer{A}}{}\Lrate{\preNtwo}{\layer{A}}{}}
 - \left((\Ra{\layer{B}})^2 
	   + 2\Ra{\layer{B}} \Rb{\layer{A}}{\layer{B}} \meanrate{\layer{B}} 
       + (\N{\layer{A}}{\layer{B}} \Rb{\layer{A}}{\layer{B}} \meanrate{\layer{B}} )^2 
   \right)										                              \notag \\
 &=
   \RbSq{\layer{A}}{\layer{B}}
   \left( 
	   \expect{\sum\limits_{\preN} \sum\limits_{\preNtwo}
				   \Lrate{\preN}{\layer{A}}{}\Lrate{\preNtwo}{\layer{A}}{}}
	 - (\N{\layer{A}}{\layer{B}} \meanrate{\layer{B}})^2
   \right). 
\end{align}
Layer $\layer{A}$ neurons are uncorrelated so that the only non-zero contribution 
to this sum occurs when a layer $\layer{A}$ neuron has a synaptic connection to each 
of the layer $\layer{B}$ neurons under consideration. In this case the input rates 
are fully correlated, so that the contribution to covariance is proportional to the 
layer $\layer{A}$ firing rate.

   \section{Derivation of radial eigenfunctions}\label{app:radialEigen}
	  \subsection{Simplified learning equation}\label{app:radialEigen_simplified}
  	     We start by decomposing the weight functions, $\Lwgtcont{\posr}{\posth}$, of 
\eq{\ref{eq:eig_k2=0_v1}}, into a sum of independent components that are dense 
in the space using a Fourier series. Therefore, expressing the exponential 
in its infinite series form, we get 
\begin{align}\label{eq:eig_k2=0_v2}
    \etalayer{} 
	\Sum{\fouriersum}{0}{\infty} \eigvalue{\fouriersum}{}
	\left( 
       \coscoeff{\fouriersum}{\posr} \cosbr{\fouriersum \posth}{} 
     + \sincoeff{\fouriersum}{\posr} \sinbr{\fouriersum \posth}  
	\right)
 \,=\, 
    \cxcoeff \radiusvar{\inputlayer}{\layer{B}} 
&  \exp{- \frac{\posr^2}{2} 
		   \left( \frac{2\radiusvar{\inputlayer}{\layer{B}} 
		               + \radiusvar{\layer{B}}{\layer{C}} }
				       { \radiusvar{\layer{B}}{\layer{C}}} \right)}
    \int_{0}^{\infty} d\posintr \, \posintr \, 
    \exp{- \frac{\posintr^2} {2}
	  	   \left( \frac{\radiusvar{\inputlayer}{\layer{B}} 
		              + \radiusvar{\layer{B}}{\layer{C}}} 
		               {\radiusvar{\layer{B}}{\layer{C}}} \right)}        \notag \\
    \int_{0}^{2\pi} d\postwoth	                                          
 &	\Sum{\expsum}{0}{\infty}
		 \left( \posr \posintr \cosbr{\posth-\postwoth}{} \right)^{\expsum}
		 \frac{1} {\expsum !}
	\Sum{\fouriersumint}{0}{\infty}
		 \left( \coscoeff{\fouriersumint}{\posintr} \cosbr{\fouriersumint \postwoth}{}
		      + \sincoeff{\fouriersumint}{\posintr} \sinbr{\fouriersumint \postwoth}{}
		 \right).
\end{align}
Rearrange the sum and integral terms to give 
\begin{align}\label{eq:eig_k2=0_v3}
    \etalayer{} 
	\Sum{\fouriersum}{0}{\infty} \eigvalue{\fouriersum}{}
    \left( 
       \coscoeff{\fouriersum}{\posr} \cosbr{\fouriersum \posth}{} 
     + \sincoeff{\fouriersum}{\posr} \sinbr{\fouriersum \posth}  
	\right)			
  \, = \, 
  & \cxcoeff \radiusvar{\inputlayer}{\layer{B}}
	\exp{- \frac{\posr^2}{2} 
		   \left( \frac{2\radiusvar{\inputlayer}{\layer{B}} + \radiusvar{\layer{B}}{\layer{C}} }
				       { \radiusvar{\layer{B}}{\layer{C}}} \right)}
    \int_{0}^{\infty} d\posintr \, \posintr \, 
    \exp{- \frac{\posintr^2} {2}
	  	   \left( \frac{2\radiusvar{\inputlayer}{\layer{B}} + \radiusvar{\layer{B}}{\layer{C}}} 
		               { \radiusvar{\layer{B}}{\layer{C}}} 
		   \right)}  \notag \\
  & \Sum{\fouriersumint}{0}{\infty} \mediumspace 
	\Sum{\expsum}{0}{\infty}
		\frac{1} {\expsum !}
		\left( \posr \posintr \right)^{\expsum}  
    \int_{0}^{2\pi} d\postwoth
	  \cosbr{\posth-\postwoth}{\expsum}
	  \left( \coscoeff{\fouriersumint}{\posintr} \cosbr{\fouriersumint \postwoth}{}
		   + \sincoeff{\fouriersumint}{\posintr} \sinbr{\fouriersumint \postwoth}
	  \right).
\end{align}

Now we consider just the inner integral over \postwoth{},
\begin{align}\label{eq:phase_int_BC}
  \int_{0}^{2\pi} d\postwoth
  \left( \cosbr{\posth-\postwoth}{\expsum} \right)
  \left( \coscoeff{\fouriersumint}{\posintr} \cosbr{\fouriersumint \postwoth}{}
	   + \sincoeff{\fouriersumint}{\posintr} \sinbr{\fouriersumint \postwoth}
  \right).
\end{align}
  
A general expression for $\left( \cosbr{\phi}{\expsum} \right)$ 
can be found by writing it as 
\begin{alignat}{2}
   \cos^{\expsum}\left( \phi \right)
=& 
   \frac{1}{2^{\expsum}}
   \left( e^{-\ii \phi} + e^{\ii \phi} \right)^{\expsum}  \notag \\
=& 
   \frac{1}{2^{\expsum}}
   \Sum{\cossum}{0}{\expsum} \binom{\expsum}{\cossum}
       \left( e^{-\ii \phi} \right)^{\cossum}
       \left( e^{ \ii \phi} \right)^{\expsum - \cossum}
  =
   \frac{1}{2^{\expsum}}
   \Sum{\cossum}{0}{\expsum} \binom{\expsum}{\cossum}
       \left( e^{-\ii \phi} \right)^{\expsum - \cossum}
       \left( e^{ \ii \phi} \right)^{\cossum}			  \notag \\
=& 
   \frac{1}{2^{\expsum}}
   \Sum{\cossum}{0}{\expsum} \binom{\expsum}{\cossum}
   \left( e^{-\ii \phi (\expsum - 2\cossum)} \right)	 
  =
   \frac{1}{2^{\expsum}}
   \Sum{\cossum}{0}{\expsum} \binom{\expsum}{\cossum}
   \left(+ e^{ \ii \phi (\expsum - 2\cossum)} \right)	  \notag \\
=& 
   \frac{1}{2^{\expsum+1}}
   \Sum{\cossum}{0}{\expsum} \binom{\expsum}{\cossum}
   \left( e^{-\ii \phi (\expsum - 2\cossum)}
        + e^{ \ii \phi (\expsum - 2\cossum)} \right)	  \notag \\
=&
   \frac{1}{2^{\expsum}}
   \Sum{\cossum}{0}{\expsum} \binom{\expsum}{\cossum}
   \cosbr{\phi(\expsum-2\cossum)}{}                       \notag \\
=& \begin{cases}
   \frac{1}{2^{\expsum-1}}
   \Sum{\cossum}{0}{\frac{\expsum-1}{2}} \binom{\expsum}{\cossum}
   \cosbr{\phi(\expsum-2\cossum)}{},	&\text{for $n$ odd}		 \\
   \frac{1}{2^{\expsum}} \binom{\expsum}{\frac{\expsum}{2}}
 + \frac{1}{2^{\expsum-1}}
   \Sum{\cossum}{0}{\frac{\expsum}{2}-1} \binom{\expsum}{\cossum}
   \cosbr{\phi(\expsum-2\cossum)}{},	&\text{for $n$ even}	 \\
   \end{cases} 
 \end{alignat}

Application of this result to \eq{\ref{eq:phase_int_BC}} gives 
\begin{align}\label{eq:phase_int_BC2}
  \int_{0}^{2\pi} d\postwoth
 &\cosbr{\posth-\postwoth}{\expsum} 
  \left( \coscoeff{\fouriersumint}{\posintr} \cosbr{\fouriersumint \postwoth}{}
	   + \sincoeff{\fouriersumint}{\posintr} \sinbr{\fouriersumint \postwoth}
  \right)															\notag \\
 =& 
  \frac{1}{2^{\expsum}}
  \int_{0}^{2\pi} d\postwoth 
  \Sum{\cossum}{0}{\expsum} \binom{\expsum}{\cossum}
       \cosbr{(\posth-\postwoth) (\expsum-2\cossum)}{}
 \left(\coscoeff{\fouriersumint}{\posintr} \cosbr{\fouriersumint \postwoth}{}
	  +\sincoeff{\fouriersumint}{\posintr} \sinbr{\fouriersumint \postwoth} \right)  \notag \\
 =& 
  \frac{1}{2^{\expsum}}
  \Sum{\cossum}{0}{\expsum} \binom{\expsum}{\cossum}
  \int_{0}^{2\pi} d\postwoth \cosbr{(\posth-\postwoth) (\expsum-2\cossum)}{}
      \coscoeff{\fouriersumint}{\posintr} \cosbr{\fouriersumint \postwoth}{}           \notag \\
 &+  
  \frac{1}{2^{\expsum}}
  \Sum{\cossum}{0}{\expsum} \binom{\expsum}{\cossum}
  \int_{0}^{2\pi} d\postwoth \cosbr{(\posth-\postwoth) (\expsum-2\cossum)}{} 
	  \sincoeff{\fouriersumint}{\posintr} \sinbr{\fouriersumint \postwoth}			\notag \\
 =& 
  \frac{1}{2^{\expsum+1}}
  \Sum{\cossum}{0}{\expsum} \binom{\expsum}{\cossum}
  \int_{0}^{2\pi} d\postwoth \coscoeff{\fouriersumint}{\posintr} 
	  \left( \cosbr{(\posth-\postwoth) (\expsum-2\cossum) 
		   + \fouriersumint \postwoth}{} 
		   + \cosbr{-(\posth-\postwoth) (\expsum-2\cossum) 
		  			+ \fouriersumint \postwoth}{}    							    
	  \right)														\notag \\
 &+   
  \frac{1}{2^{\expsum+1}}
  \Sum{\cossum}{0}{\expsum} \binom{\expsum}{\cossum}
  \int_{0}^{2\pi} d\postwoth \sincoeff{\fouriersumint}{\posintr} 
	  \left( \sinbr{ (\posth-\postwoth) (\expsum-2\cossum) 
		            + \fouriersumint \postwoth} 
	       - \sinbr{-(\posth-\postwoth) (\expsum-2\cossum) 
		            + \fouriersumint \postwoth} 
	  \right)														\notag \\
 =&
  \frac{1}{2^{\expsum+1}}
  \Sum{\cossum}{0}{\expsum} \binom{\expsum}{\cossum}
  \int_{0}^{2\pi} d\postwoth \coscoeff{\fouriersumint}{\posintr} 
	  \left( \cosbr{\posth(\expsum-2\cossum) 
			      + \postwoth(\fouriersumint - (\expsum-2\cossum))}{} 
		   + \cosbr{\posth(\expsum-2\cossum) 
			      + \postwoth(\fouriersumint + (\expsum-2\cossum))}{}    							    
	  \right)														\notag \\
 &+   
  \frac{1}{2^{\expsum+1}}
  \Sum{\cossum}{0}{\expsum} \binom{\expsum}{\cossum}
  \int_{0}^{2\pi} d\postwoth \sincoeff{\fouriersumint}{\posintr} 
	  \left( \sinbr{ \posth(\expsum-2\cossum) 
			      + \postwoth(\fouriersumint - (\expsum-2\cossum))} 
	       - \sinbr{\posth(\expsum-2\cossum) 
			      + \postwoth(\fouriersumint + (\expsum-2\cossum))} 
	  \right)														\notag \\
=& \begin{dcases*}
   \frac{1}{2^{\expsum}}
   \Sum{\cossum}{0}{\frac{\expsum-1}{2}} \binom{\expsum}{\cossum}
   \int_{0}^{2\pi} d\postwoth 
   \left[ \coscoeff{\fouriersumint}{\posintr} 
	  \left( \cosbr{\posth(\expsum-2\cossum) 
			      + \postwoth(\fouriersumint - (\expsum-2\cossum))}{} 
		   + \cosbr{\posth(\expsum-2\cossum) 
			      + \postwoth(\fouriersumint + (\expsum-2\cossum))}{}    							    
	  \right) \right.														  	  \\
      \quad  + \left. \sincoeff{\fouriersumint}{\posintr} 
	  \left( \sinbr{\posth(\expsum-2\cossum) 
			      + \postwoth(\fouriersumint - (\expsum-2\cossum))}{} 
		   + \sinbr{\posth(\expsum-2\cossum) 
			      - \postwoth(\fouriersumint + (\expsum-2\cossum))}{}    							    
	  \right)
   \right],	 & for $n$ odd             		                                      \\
   \frac{1}{2^{\expsum+1}} \binom{\expsum}{\frac{\expsum}{2}} 
   \int_{0}^{2\pi} d\postwoth 
   \left( \cosbr{\fouriersumint \postwoth}{} 
		  + \sinbr{\fouriersumint \postwoth}{}  \right)                           \\
   \quad  + \frac{1}{2^{\expsum}}
   \Sum{\cossum}{0}{\frac{\expsum}{2}-1} \binom{\expsum}{\cossum} 
   \int_{0}^{2\pi} d\postwoth 
   \left[ \coscoeff{\fouriersumint}{\posintr} 
	  \left( \cosbr{\posth(\expsum-2\cossum) 
			      + \postwoth(\fouriersumint - (\expsum-2\cossum))}{} 
		   + \cosbr{\posth(\expsum-2\cossum) 
			      + \postwoth(\fouriersumint + (\expsum-2\cossum))}{}    							    
	  \right) \right.															  \\
      \quad  + \sincoeff{\fouriersumint}{\posintr} 
	  \left. \left( \sinbr{\posth(\expsum-2\cossum) 
			      + \postwoth(\fouriersumint - (\expsum-2\cossum))}{} 
		   + \sinbr{\posth(\expsum-2\cossum) 
			      - \postwoth(\fouriersumint + (\expsum-2\cossum))}{}    							    
	  \right)
   \right],	& for $n$ even                                                        \\
   \end{dcases*} 
\end{align}
All of the integrals within the binomial sum term will evaluate to zero since the 
functions are periodic in $2\pi$ and centred around a mean of zero, except those 
for which $\fouriersumint = \expsum-2\cossum$, because in these cases the $\postwoth$ 
terms cancel, and therefore the integration is over a constant. For odd \fouriersumint, 
this can only happen for odd \expsum, and for even \fouriersumint, this can only 
happen for even \expsum, when $\cossum = \frac{\expsum - \fouriersumint}{2}$  
such that $0 \leq \cossum \leq \frac{\expsum-1}{2}$. Consequently, the sinusoidal 
term for which the \postwoth{} coefficient is $\fouriersumint + \expsum - 2\cossum$ 
will always integrate to $0$, since $\fouriersumint + 2\cossum-\expsum \ge 0$ for 
all \cossum{}. The additional term for even $\expsum$ will only be non-zero 
when $\fouriersumint$ is $0$.  Thus equation \eq{\ref{eq:phase_int_BC2}} evaluates to 
\begin{align}\label{eq:cosineInt}
  \int_{0}^{2\pi} d\postwoth
  \cosbr{\posth-\postwoth}{\expsum} 
 &\left( \coscoeff{\fouriersumint}{\posintr} \cosbr{\fouriersumint \postwoth}{}
	   + \sincoeff{\fouriersumint}{\posintr} \sinbr{\fouriersumint \postwoth}
  \right)																		\notag \\
 =& 
 \begin{cases}
    \frac{2\pi}{2^{\expsum+1}} \binom{\expsum}{\frac{\expsum}{2}}
    \coscoeff{\fouriersumint}{\posintr}, 
  & \text{for } \fouriersumint{}=0 \text{ and } \expsum \text{ even}                   \\
    \frac{2\pi}{2^{\expsum}}
    \binom{\expsum}{\frac{\expsum-\fouriersumint}{2}}
    \coscoeff{\fouriersumint}{\posintr} \cosbr{\fouriersumint \posth}{} 
  +    
    \frac{2\pi}{2^{\expsum}}
    \binom{\expsum}{\frac{\expsum-\fouriersumint}{2}}
    \sincoeff{\fouriersumint}{\posintr} \sinbr{\fouriersumint \posth}, 
  & \text{for } 0 \leq \frac{\expsum-\fouriersumint}{2} \leq \frac{\expsum-1}{2} 
    \text{ an integer} 			                                                       \\
    0,
  & \text{otherwise}.
 \end{cases}
\end{align}

Note that for $\expsum - \fouriersumint \geq 0$ we require 
$\expsum \geq \fouriersumint$, and thus the infinite sum over \fouriersumint{} in 
\eq{\ref{eq:eig_k2=0_v3}} can be truncated. 

Incorporating this result into the infinite sums from \eq{\ref{eq:eig_k2=0_v3}}, 
separating the odd and even terms for \expsum{} and \fouriersumint{}, and focusing 
only on the cosine components for the interim, gives, \newline 
\begin{align}
   \Sum{\fouriersumint}{0}{\infty} \mediumspace 
   \Sum{\expsum}{0}{\infty}
 & \int_{0}^{2\pi} d\postwoth
	 \cosbr{\posth-\postwoth}{\expsum}
	 \coscoeff{\fouriersumint}{\posintr} \cosbr{\fouriersumint \postwoth}{}  \notag \\
 &= 
   \Sum{\expsum}{0}{\infty}
   \Sum{\fouriersumint}{0}{\expsum} \mediumspace 
   \left[
	   \frac{2\pi}{2^{2\expsum}}
	   \binom{2\expsum}{\expsum-\fouriersumint}
	   \coscoeff{2\fouriersumint}{\posintr} \cosbr{2\fouriersumint \posth}{} 
   + 
	   \frac{2\pi}{2^{2\expsum+1}}
	   \binom{2\expsum+1}{\expsum-\fouriersumint}
	   \coscoeff{2\fouriersumint+1}{\posintr} \cosbr{(2\fouriersumint+1) \posth}{} 
   \right]																\notag 	  \\
 &= 
   \Sum{\fouriersumint}{0}{\infty} \mediumspace 
   \Sum{\expsum}{\fouriersumint}{\infty}
   \left[
	   \frac{2\pi}{2^{2\expsum}}
	   \binom{2\expsum}{\expsum-\fouriersumint}
	   \coscoeff{2\fouriersumint}{\posintr} \cosbr{2\fouriersumint \posth}{} 
   + 
	   \frac{2\pi}{2^{2\expsum+1}}
	   \binom{2\expsum+1}{\expsum-\fouriersumint}
	   \coscoeff{2\fouriersumint+1}{\posintr} \cosbr{(2\fouriersumint+1) \posth}{} 
   \right]. 
\end{align}
Let $\expsumtwo = \expsum - \fouriersumint$, so that 
\begin{align}
   \Sum{\fouriersumint}{0}{\infty} \mediumspace 
   \Sum{\expsum}{0}{\infty}
 & \int_{0}^{2\pi} d\postwoth
	 \cosbr{\posth-\postwoth}{\expsum}
	 \coscoeff{\fouriersumint}{\posintr} \cosbr{\fouriersumint \postwoth}{} \notag  \\
 &= 
   \Sum{\fouriersumint}{0}{\infty} \mediumspace 
   \Sum{\expsumtwo}{0}{\infty}
   \left[
	   \frac{2\pi}{2^{2(\expsumtwo + \fouriersumint)}}
	   \binom{2(\expsumtwo + \fouriersumint)}{\expsumtwo}
	   \coscoeff{2\fouriersumint}{\posintr} \cosbr{2\fouriersumint \posth}{} 
   + 
	   \frac{2\pi}{2^{2(\expsumtwo + \fouriersumint)+1}}
	   \binom{2(\expsumtwo + \fouriersumint) +1 }{\expsumtwo}
	   \coscoeff{2\fouriersumint+1}{\posintr} \cosbr{(2\fouriersumint+1) \posth}{} 
   \right]. 
\end{align}
Note that \expsumtwo{} and \fouriersumint{} can now be set independently so that 
the odd and even terms for \fouriersumint{} can now be recombined. 
\begin{align}
   \Sum{\fouriersumint}{0}{\infty} \mediumspace 
   \Sum{\expsumtwo}{0}{\infty}
   \int_{0}^{2\pi} d\postwoth
	 \cosbr{\posth-\postwoth}{\expsum}
	 \coscoeff{\fouriersumint}{\posintr} \cosbr{\fouriersumint \postwoth}{} 
 &= 
   \Sum{\fouriersumint}{0}{\infty} \mediumspace 
   \Sum{\expsumtwo}{0}{\infty}
   \frac{2\pi}{2^{2\expsumtwo + \fouriersumint}}
   \binom{2\expsumtwo + \fouriersumint}{\expsumtwo}
   \coscoeff{\fouriersumint}{\posintr} \cosbr{\fouriersumint \posth}{}.
\end{align}

Incorporating this sum into the full equation for \eq{\ref{eq:eig_k2=0_v3}}, gives, 
\begin{align}\label{eq:eig_k2=0_simplifiedTheta}
    \etalayer{} 
	\Sum{\fouriersum}{0}{\infty} \eigvalue{\fouriersum}{}
	\left( 
       \coscoeff{\fouriersum}{\posr} \cosbr{\fouriersum \posth}{} \right.
  & \left.
     + \sincoeff{\fouriersum}{\posr} \sinbr{\fouriersum \posth}  
	\right)                                                                              \notag \\
  \mediumspace = \mediumspace
  & \cxcoeff \radiusvar{\inputlayer}{\layer{B}} 
	\exp{- \frac{\posr^2}{2} 
		   \left( \frac{2\radiusvar{\inputlayer}{\layer{B}} + \radiusvar{\layer{B}}{\layer{C}} }
				       { \radiusvar{\layer{B}}{\layer{C}}} \right)}
    \int_{0}^{\infty} d\posintr \, \posintr \, 
    \exp{- \frac{\posintr^2} {2}
	  	   \left( \frac{2\radiusvar{\inputlayer}{\layer{B}} + \radiusvar{\layer{B}}{\layer{C}}} 
		               { \radiusvar{\layer{B}}{\layer{C}}} \right)}                                  \notag \\
  & \Sum{\fouriersumint}{0}{\infty} \mediumspace 
	\Sum{\expsumtwo}{0}{\infty}
		\binom{2\expsumtwo+\fouriersumint}{\expsumtwo}
		\frac{1} {(2\expsumtwo+\fouriersumint) !}
		\left( \posr \posintr \right)^{2\expsumtwo+\fouriersumint}  
		\frac{2\pi}{2^{2\expsumtwo+\fouriersumint}}
		\left( \coscoeff{\fouriersumint}{\posintr} \cosbr{\fouriersumint \posth}{} 
	         + \sincoeff{\fouriersumint}{\posintr} \sinbr{\fouriersumint \posth} \right) \notag \\
  \mediumspace = \mediumspace
  & \cxcoeff \radiusvar{\inputlayer}{\layer{B}}
	\exp{- \frac{\posr^2}{2} 
		   \left( \frac{2\radiusvar{\inputlayer}{\layer{B}} + \radiusvar{\layer{B}}{\layer{C}} }
				       { \radiusvar{\layer{B}}{\layer{C}}} \right)}
    \int_{0}^{\infty} d\posintr \posintr 
    \exp{- \frac{\posintr^2} {2}
	  	   \left( \frac{2\radiusvar{\inputlayer}{\layer{B}} + \radiusvar{\layer{B}}{\layer{C}}} 
		               { \radiusvar{\layer{B}}{\layer{C}}} \right)}                                  \notag \\
  & \Sum{\fouriersumint}{0}{\infty} \mediumspace 
	\Sum{\expsumtwo}{0}{\infty}
		\frac{2\pi} {\expsumtwo ! (\expsumtwo+\fouriersumint)!}
		\left( \frac{ \posr \posintr } {2} \right)^{2\expsumtwo+\fouriersumint}  
		\left( \coscoeff{\fouriersumint}{\posintr} \cosbr{\fouriersumint \posth}{} 
	         + \sincoeff{\fouriersumint}{\posintr} \sinbr{\fouriersumint \posth} \right) \notag \\
  \mediumspace = \mediumspace
  & 2\pi \cxcoeff \radiusvar{\inputlayer}{\layer{B}} 
	\exp{- \frac{\posr^2}{2} 
		   \left( \frac{2\radiusvar{\inputlayer}{\layer{B}} + \radiusvar{\layer{B}}{\layer{C}} }
				       { \radiusvar{\layer{B}}{\layer{C}}} \right)}
    \int_{0}^{\infty} d\posintr \posintr 
    \exp{- \frac{\posintr^2} {2}
	  	   \left( \frac{2\radiusvar{\inputlayer}{\layer{B}} + \radiusvar{\layer{B}}{\layer{C}}} 
		               { \radiusvar{\layer{B}}{\layer{C}}} \right)}                                 \notag \\
  & \Sum{\fouriersumint}{0}{\infty} \mediumspace 
	\Sum{\expsumtwo}{0}{\infty}
		\frac{1} 
             {\expsumtwo ! \Gamma(\expsumtwo+\fouriersumint+1)}
		\left( \frac{\posr \posintr } {2} \right)^{2\expsumtwo+\fouriersumint}  
		\left( \coscoeff{\fouriersumint}{\posintr} \cosbr{\fouriersumint \posth}{} 
	         + \sincoeff{\fouriersumint}{\posintr} \sinbr{\fouriersumint \posth} 
		\right)                                                                        \notag \\
  \mediumspace = \mediumspace
  & 2\pi \cxcoeff \radiusvar{\inputlayer}{\layer{B}} 
	\exp{- \frac{\posr^2}{2} 
		   \left( \frac{2\radiusvar{\inputlayer}{\layer{B}} 
				       + \radiusvar{\layer{B}}{\layer{C}} }
				       { \radiusvar{\layer{B}}{\layer{C}}} \right) }								\notag \\
  &\phantom{=2\pi A}
    \Sum{\fouriersumint}{0}{\infty} \mediumspace 
   	    \int_{0}^{\infty} d\posintr \posintr                                       
			\exp{-\frac{\posintr^2}{2}
				  \left( \frac{2\radiusvar{\inputlayer}{\layer{B}} 
						      + \radiusvar{\layer{B}}{\layer{C}} }
							  { \radiusvar{\layer{B}}{\layer{C}} }
				  \right)} 
			\bessel{\fouriersumint}{ \posr \posintr }  
			\left( \coscoeff{\fouriersumint}{\posintr} \cosbr{\fouriersumint \posth}{} 
				 + \sincoeff{\fouriersumint}{\posintr} \sinbr{\fouriersumint \posth} \right), 
\end{align}
where $\bessel{\alpha}{x}$ is a modified Bessel function of the first kind, of 
order $\alpha$, such that 
\begin{align}\label{eq:bessel}
   \bessel{\alpha}{x} = \Sum{m}{0}{\infty}  \frac{1}{ m!\Gamma (m+\alpha+1) }
                            \left( \frac{x}{2} \right)^{2m+\alpha} . 
\end{align}

We can consider each component in the sum separately, such that  
\begin{align}\label{eq:eigBC_integral_v3}
    \etalayer{} \eigvalue{\fouriersum}{}
 &  \left( \coscoeff{\fouriersum}{\posr} \cosbr{\fouriersum \posth}{} 
         + \sincoeff{\fouriersum}{\posr} \sinbr{\fouriersum \posth}  
    \right)													  \notag \\
   \mediumspace = \mediumspace
 & 2\pi \cxcoeff \radiusvar{\inputlayer}{\layer{B}} 
	\exp{- \frac{\posr^2}{2} 
		   \left( \frac{2\radiusvar{\inputlayer}{\layer{B}} + \radiusvar{\layer{B}}{\layer{C}} }
				       { \radiusvar{\layer{B}}{\layer{C}}} \right)}
    \int_{0}^{\infty} d\posintr \posintr 
    \exp{-\frac{\posintr^2}{2}
	  	  \left( \frac{2\radiusvar{\inputlayer}{\layer{B}} + \radiusvar{\layer{B}}{\layer{C}} }
					  { \radiusvar{\layer{B}}{\layer{C}} }
		  \right)} 
        \bessel{\fouriersumint}{ \posr \posintr }
     	\left( \coscoeff{\fouriersumint}{\posintr} 
			   \cosbr{\fouriersumint \posth}{} 
	         + \sincoeff{\fouriersumint}{\posintr} 
			   \sinbr{\fouriersumint \posth} \right).
\end{align}

To derive the eigenfunctions that satisfy \eq{\ref{eq:eigBC_integral_v3}}, 
we require weight functions with an exponential of the same form, and polynomials 
in $\posr$ that will be of the same order after evaluating the integral. Furthermore, 
it is well known that Laguerre polynomials are orthogonal over the interval 
$[0,\infty)$, with respect to the weight function $x^a\exp{-x}$. 
Consequently, we propose eigenfunctions of the form 
\begin{align}\label{eq:angularWeightFunction}
   \coscoeff{\fouriersum,\eigLindex}{\posintr} 
   \begin{cases}
       \cosbr{(\fouriersumint-\eigLindex) \posth}{}  \\
       \sinbr{(\fouriersumint-\eigLindex) \posth}{}
   \end{cases}
  = \largespace
   \eignorm{\fouriersum}{\eigLindex} 
   \posintr^{\fouriersum-\eigLindex} \exp{-\frac{\posintr^2}{2 \eigdecay}} 
   \laguerre{\eigLindex}{\fouriersum-\eigLindex}{\frac{\posintr^2}{\eigdecay}}
   \begin{cases}
       \cosbr{(\fouriersumint-\eigLindex) \posth}{}  \\
       \sinbr{(\fouriersumint-\eigLindex) \posth}{}
   \end{cases},
\end{align}
where the additional index, $\eigLindex$, denotes the index into eigenvalues of the 
same order, $\fouriersum$, since the solutions are degenerate, and 
$\eignorm{\fouriersum}{\eigLindex}$ is a normalisation factor. 
$\laguerre{\eigLindex}{\fouriersum-\eigLindex}{}$ is an associated Laguerre polynomial. 
Since $\int_0^\infty x^p e^{-x}\mathrm{L}_q^p(x)^2\,dx=(p+q)!/q!$, the normalisation 
factor can be derived as,
\begin{align}
    \eignorm{\fouriersum}{\eigLindex} 
 &= 
    \begin{cases}
		\sqrt{ \frac{2 \eigLindex!}
					{\pi \fouriersum! \eigdecay
					 \radiusvar{\inputlayer}{\layer{B}} } }, 
		\qquad &\fouriersum=\eigLindex \\
		\sqrt{ \frac{\eigLindex!}
					{\pi \fouriersum! \eigdecay^{\fouriersum-\eigLindex+1}
					 \radiusvar{\inputlayer}{\layer{B}} } },
		\qquad &\textrm{otherwise},
	\end{cases}
\end{align}
where the factor of $2$ difference occurs for the case $\fouriersum=\eigLindex$, 
because the integral for the angular component is over $\cos{(0\posth)}$, a constant. 

Separating the $\cos$ and $\sin$ terms in \eq{\ref{eq:eigBC_integral_v3}} since 
they are independent components, and letting
\begin{align}\label{eq:expcoeff}
   \expcoeff 
 &= 
   \frac{ \radiusvar{\layer{B}}{\layer{C}} }
        {2\radiusvar{\inputlayer}{\layer{B}} 
	    + \radiusvar{\layer{B}}{\layer{C}} }
\end{align}
the eigenfunctions must satisfy 
\begin{align}\label{eq:eigBC_integral_v4}
    \posr ^{\fouriersum-\eigLindex} 
	\exp{-\frac{\posr^2}{2\eigdecay}} 
 &  \laguerre{\eigLindex}{\fouriersum-\eigLindex}{\frac{\posr^2}{\eigdecay}}
    \cosbr{(\fouriersum-\eigLindex) \posth}{}						\notag \\
 &\overset{?}{=}
   2\pi \cxcoeff \radiusvar{\inputlayer}{\layer{B}} 
	\exp{- \frac{\posr^2}{2\expcoeff} }
    \int_{0}^{\infty} d\posintr \posintr 
        \exp{-\frac{\posintr^2}{2\expcoeff}}
		\exp{ \frac{- \posintr^2}{2\eigdecay}}
     	\posintr^{\fouriersum-\eigLindex} 
        \laguerre{\eigLindex}{\fouriersum-\eigLindex}{\frac{\posintr^2}{\eigdecay}}
		\bessel{\fouriersum-\eigLindex}{ \posr \posintr }	
		\cosbr{(\fouriersum-\eigLindex) \posth}{}                  \notag \\
 &=
   2\pi \cxcoeff \radiusvar{\inputlayer}{\layer{B}}
	\exp{- \frac{\posr^2}{2\expcoeff}}
    \int_{0}^{\infty} d\posintr \posintr 
       \exp{-\frac{\posintr^2}{2} 
		     \left( \frac{\expcoeff + \eigdecay}{\expcoeff \eigdecay} \right) }
     	\posintr^{\fouriersum-\eigLindex} 
		\bessel{\fouriersum-\eigLindex}{ \posr \posintr }
        \laguerre{\eigLindex}{\fouriersum-\eigLindex}{\frac{\posintr^2}{\eigdecay}}
		\cosbr{(\fouriersum-\eigLindex) \posth}{}.
\end{align}
Note the following integral, evaluated using \citet{Mathematica}:
\begin{align}\label{eq:general_integral}
    J_{\eigLindex,\fouriersum} 
 &= 
    \int_{0}^{\infty} d\posintr \posintr 
	   \exp{-\frac{\posintr^2}{2B}} \posintr^{\fouriersum-\eigLindex} 
	   \bessel{\fouriersum-\eigLindex}{\posr \posintr}
	   \laguerre{ \eigLindex }{ \fouriersum - \eigLindex }{ \frac{\posintr^2}{\eigdecay} } \notag \\
 &= 
    B^{\fouriersum+1} \left( \frac{ \eigdecay - 2B }{ B\eigdecay} \right)^{\eigLindex}
	\exp{\frac{B\posr^2}{2}} \posr^{\fouriersum - \eigLindex} 
	\laguerre{\eigLindex}{\fouriersum-\eigLindex}{\frac{B^2\posr^2}{\eigdecay - 2B}}. 
\end{align}
Applying this integral to \eq{\ref{eq:eigBC_integral_v4}} gives,
\begin{align}
    \posr ^{\fouriersum-\eigLindex} 
 &  \exp{-\frac{\posr^2}{2\eigdecay}} 
    \laguerre{\eigLindex}{\fouriersum-\eigLindex}{\frac{\posr^2}{\eigdecay}}
    \cosbr{(\fouriersum-\eigLindex) \posth}{}						\notag \\
 &\overset{?}{=}
   2\pi \cxcoeff \radiusvar{\inputlayer}{\layer{B}} 
	\exp{- \frac{\posr^2}{2\expcoeff}}
    \left( \frac{\expcoeff \eigdecay}{\expcoeff + \eigdecay} \right)^{\fouriersum+1}
	\left( \frac{ \eigdecay - 2\left( \frac{\expcoeff + \eigdecay}{\expcoeff \eigdecay} \right) }
			    { \left( \frac{\expcoeff + \eigdecay}{\expcoeff \eigdecay} \right) \eigdecay} 
				  \right)^{\eigLindex}
	\posr^{\fouriersum - \eigLindex} 
	\exp{\left( \frac{\expcoeff \eigdecay}{\expcoeff + \eigdecay} \right) \frac{\posr^2}{2}} 
	\laguerre{\eigLindex}{\fouriersum-\eigLindex}
             {\frac{\left( \frac{\expcoeff + \eigdecay}{\expcoeff \eigdecay} \right)^2\posr^2}
				   {\eigdecay - 2\left( \frac{\expcoeff + \eigdecay}{\expcoeff \eigdecay} \right)}} 
    \cosbr{(\fouriersum-\eigLindex) \posth}{}						\notag \\
 &= 
  2\pi \cxcoeff \radiusvar{\inputlayer}{\layer{B}} 
    \left( \frac{ \eigdecay \expcoeff }{\expcoeff + \eigdecay} \right)^{\fouriersum+1} 
	\left( \frac{ \eigdecay - \expcoeff }{ \eigdecay + \expcoeff } \right) ^{\eigLindex}
	\posr^{\fouriersum - \eigLindex} 
	\exp{- \left( \frac{\expcoeff + \eigdecay - 2\expcoeff^2\eigdecay}
		 	           {\expcoeff (\expcoeff + \eigdecay)}
		   \right) \frac{\posr^2}{2}} 
	\laguerre{\eigLindex}{\fouriersum-\eigLindex}
             {\frac{ \eigdecay \expcoeff^2 \posr^2 }
				   { \eigdecay^2 - \expcoeff^2} } 
    \cosbr{(\fouriersum-\eigLindex) \posth}{}.						\notag \\
\end{align}
For this equivalence to be true, it is necessary to equate terms. After some 
simplification, it can be seen that equating both the exponential and Laguerre 
terms requires that,
\begin{align}
   \frac{1}{\eigdecay} &= \frac{\eigdecay \expcoeff^2} {\eigdecay^2 - \expcoeff^2}  \notag \\
 \rightarrow
                     0 &= \eigdecay (\expcoeff^2 - 1) + \expcoeff^2.
\end{align}
Solving this quadratic in $\eigdecay{}$ requires,
\begin{align}
   \eigdecay 
 &= 
   \frac{\expcoeff}{ \sqrt{1 - \expcoeff^2 }},
\end{align}
from which condition we finally get,
\begin{align}\label{eq:eigBC_integral_v5}
    \posr ^{\fouriersum-\eigLindex} \exp{-\frac{\posr^2}{2\eigdecay}} 
 &  \laguerre{\eigLindex}{\fouriersum-\eigLindex}{\frac{\posr^2}{\eigdecay}}
    \cosbr{(\fouriersum-\eigLindex) \posth}{}						\notag \\
 &= 
  2\pi \cxcoeff \radiusvar{\inputlayer}{\layer{B}}
    \left( \frac{ \eigdecay \expcoeff }{\expcoeff + \eigdecay} \right)^{\fouriersum+\eigLindex+1} 
	\posr^{\fouriersum - \eigLindex} 
	\exp{- \frac{\posr^2}{2\eigdecay} } 
	\laguerre{\eigLindex}{\fouriersum-\eigLindex}
             {\frac{\posr^2}{\eigdecay}}                               
    \cosbr{(\fouriersum-\eigLindex) \posth}{}.						\notag \\
\end{align}
Substituting the radial connection parameters back in using \eq{\ref{eq:expcoeff}} 
finally gives, 
\begin{align}\label{app:eigendecay_radial}
   \eigdecay 
 &= 
   \frac{ \radiusvar{\layer{B}}{\layer{C}} } 
        { 2 \radius{\inputlayer}{\layer{B}} 
			\sqrt{\radiusvar{\inputlayer}{\layer{B}} + \radiusvar{\layer{B}}{\layer{C}} }}. 
\end{align}

Consequently, the eigenfunctions and eigenvalues for the learning equation can be 
expressed in polar coordinates as,
\begin{subequations}\label{app:simpleEigenSoln}
\begin{align}
    \eigvalue{\fouriersum}{\eigLindex} 
 &= 
    2 \pi \cxcoeff 
    \left( \frac{\eigdecay   \radiusvar{\layer{B}}{\layer{C}} }
			    {\eigdecay ( \radiusvar{\inputlayer}{\layer{B}} 
						  + \radiusvar{\layer{B}}{\layer{C}} )
			   + \radiusvar{\layer{B}}{\layer{C}} }
	\right)^{\fouriersum+\eigLindex+1}			        \label{app:simpleEigenvalue} \\
    \eigvec{\fouriersum,\eigLindex}{\posr,\posth}
 &= 
    \eignorm{\fouriersum}{\eigLindex} \posr ^{\fouriersum-\eigLindex} 
    \exp{-\frac{\posr^2}{2\eigdecay}} 
    \laguerre{\eigLindex}{\fouriersum-\eigLindex}{\frac{\posr^2}{\eigdecay}}
	\exp{ \imag (\fouriersum-\eigLindex) \posth } \, .  \label{app:simpleEigenfn} 
\end{align}
\end{subequations}

	  \subsection{Full learning equation}\label{app:radialEigen_full}
	   
If the simplified learning equation in \eq{\ref{eq:learningEqnCont}} 
is denoted by $\operatorSimple$, from the eigenfunctions derived for the 
simplified learning equation, we know that,  
\begin{align}
    \eigvalueSimple{\fouriersum}{\eigLindex} 
	\eigvecSimple{\fouriersum,\eigLindex}{\posr,\posth}
  &= 
    \operatorSimple \eigvecSimple{\fouriersum,\eigLindex}{\posr,\posth}.
\end{align}
If we perturb the simplified learning equation by adding a small 
$\ktwo{\layer{B}}{\layer{C}}$, denote the perturbed system by $\operatorFull$, 
and the eigenfunctions of the perturbed system by 
$\eigvecFull{\fouriersum,\eigLindex}{\posr,\posth}$, where $\fouriersum$
and $\eigLindex$ determine the order of the eigenfunction, so that 
\begin{align}
    \operatorFull
 &= \cxcoeff \: 
    \int_{0}^{\infty} d\posintr \: \posintr \int_{0}^{2\pi} d\postwoth
	  \left( 
		  \exp{- \frac{ \posr^2 + \posintr^2 
									  -2\posr \posintr \cosbr{\posth-\postwoth}{} }
									  {2\radiusvar{\inputlayer}{\layer{B}} }} 
		+ \ktwo{\layer{B}}{\layer{C}} 
	  \right) 
	  \exp{- \frac{ \posintr^2 + \posr^2 }
				  { \radiusvar{\layer{B}}{\layer{C}}}}
	  \Lwgtcont{\posintr}{\postwoth}, 
\end{align}
and the eigenvectors for the full system, including the perturbation, satisfy
\begin{align}
    \eigvalueFull{\fouriersum}{\eigLindex} 
	\eigvecFull{\fouriersum,\eigLindex}{\posr,\posth}
  &= 
    \operatorFull \eigvecFull{\fouriersum,\eigLindex}{\posr,\posth}.
\end{align}

The perturbation on the integral operator, denoted \operatorPerturb, is then, 
\begin{align}\label{eq:perturbOp}
    \operatorPerturb
 &= \cxcoeff \: 
    \int_{0}^{\infty} d\posintr \: \posintr \int_{0}^{2\pi} d\postwoth \:
	  \ktwo{\layer{B}}{\layer{C}} \:
	  \exp{- \frac{ \posintr^2 + \posr^2 }
				  { \radiusvar{\layer{B}}{\layer{C}}}}
	  \eigvecSimple{\fouriersum,\eigLindex}{\posr,\posth}, 
\end{align}
and we require the new eigenfunctions to be similar to the eigenfunctions of the 
simplified learning equation, plus a small perturbation, so that the first order 
corrections to the eigenfunctions and eigenvalues can be defined as,
\begin{align}
   \eigvecFull{\fouriersum,\eigLindex}{\posr,\posth}
 &= 
   \eigvecSimple{\fouriersum,\eigLindex}{\posr,\posth}
 + \eigvecPerturb{\fouriersum,\eigLindex}{\posr,\posth},  \tab\tab
   \eigvalueFull{\fouriersum}{\eigLindex} 
  = 
   \eigvalueSimple{\fouriersum}{\eigLindex} 
 + \eigvaluePerturb{\fouriersum}{\eigLindex}. 
\end{align}

If 
$ W_{\fouriersum,\eigLindex}^{m,p} = 
  \int_{-\infty}^{\infty} d\posr \, \posr 
     \int_{0}^{2\pi} d\posth
		\conj{ \eigvecSimple{\fouriersum, \eigLindex}{\posr,\posth} }
	       \:  \operatorPerturb
	       \:  \eigvecSimple{m, p}{\posr,\posth} $, 
for non-degenerate eigenfunctions, , i.e.  $\fouriersum + \eigLindex \neq m + p$ 
the first order corrections can be determined 
by \citep{Kat95}
\begin{align}\label{eq:eigvecPerturb}
    \eigvecPerturb{\fouriersum,\eigLindex}{\posr,\posth}  
  &= 
    \sum\limits_{\fouriersum+\eigLindex \neq m+p}
	\frac{ W_{\fouriersum,\eigLindex}^{m,p} }
	     { \eigvalueSimple{\fouriersum}{\eigLindex} 
		 - \eigvalueSimple{m}{p} }, 
    \eigvecSimple{\fouriersum,\eigLindex}{\posr,\posth}  
	\largespace
	\textrm{and}
	\largespace
   \eigvaluePerturb{\fouriersum}{\eigLindex} 
  = 
   W_{\fouriersum,\eigLindex}^{l,n},
\end{align}
for non-degenerate eigenfunctions, i.e. 
$\fouriersum + \eigLindex \neq m + p$. For degenerate eigenfunctions, 
where the denominator of \eq{\ref{eq:eigvecPerturb}} is equal to zero, 
the first order correction of degenerate eigenfunctions of order 
$\fouriersum+\eigLindex$, can be found as a weighted sum of the degenerate 
eigenfunctions, where the weights are determined by the eigenvectors of the  
$(\fouriersum+\eigLindex) \times (\fouriersum+\eigLindex)$ matrix of
$W_{\fouriersum,\eigLindex}^{m,p}$ coefficients. Given that the set of degenerate 
eigenfunctions of order $\fouriersum+\eigLindex$ have angular terms with 
different frequencies the off-diagonal terms of this matrix are $0$. Hence, 
the eigenfunctions of this matrix are simply the terms,
$W_{\fouriersum,\eigLindex}^{\fouriersum,\eigLindex}$. Additionally, where 
$\fouriersum \neq \eigLindex$, the diagonal terms will be zero because the 
perturbation integrates to $0$ for each radii, which can be seen from 
the equal number of light and dark regions as you traverse from $0$ to 
$2\pi$ at a given radius in 
\fig{\ref{fig:radialEigenfunctions}}. Consequently, the only non-zero 
perturbation term in a set of degenerate eigenfunctions are those for which 
$\fouriersum = \eigLindex$, which happens once, and only where 
$\fouriersum + \eigLindex$ is even.

For a pair of non-degenerate eigenfunctions where they have different 
angular frequencies the perturbation will also evaluate to zero. However, 
note that where the values $\fouriersum-\eigLindex$ equals $m-p$, so that 
the angular terms have the same frequency, some non-zero terms can appear. 
However, these terms are very small so can be ignored for the purposes of 
the perturbation approximation. It is therefore only necessary to evaluate 
the diagonal terms, which are denoted by the single index pair, 
$W_{\fouriersum,\eigLindex}$, and can be evaluated as, 
\begin{align}
   W_{\fouriersum,\eigLindex}
 &= 
   2\pi \cxcoeff \: \eignorm{\fouriersum}{\eigLindex}^2 
   \int_{0}^{\infty} d\posintr \: \posintr \:
      \posintr^{2 (\fouriersum-\eigLindex)} \exp{-\frac{\posr^2}{\eigdecay}} 
	  \exp{- \frac{ \posintr^2 + \posr^2 }
				  { \radiusvar{\layer{B}}{\layer{C}}}}
      \left[ \laguerre{\eigLindex}{\fouriersum-\eigLindex}{\frac{\posr^2}{\eigdecay}} 
	  \right]^2  \notag \\
 &= 
   \pi \cxcoeff \: \eignorm{\fouriersum}{\eigLindex}^2 \eigdecay^{\fouriersum + \eigLindex + 1}
   \exp{- \frac{ \posr^2 }{ \radiusvar{\layer{B}}{\layer{C}}} }
   \int_{0}^{\infty} dz \:
      z^{ \fouriersum-\eigLindex } \exp{-\alpha z}
      \left[ \laguerre{\eigLindex}{\fouriersum-\eigLindex}{z} \right]^2,
\end{align}
where $\alpha = \frac{ \radiusvar{\layer{B}}{\layer{C}} + \eigdecay }
                     { \radiusvar{\layer{B}}{\layer{C}} } $. 
This integral on the right hand side can be evaluated as \citep{GraRyz07},
\begin{align}
   \int_0^{\infty} e^{-bx} x^{a} \laguerre{n}{a}{x} \laguerre{m}{a}{x} dx 
 &= 
   \frac{ \Gamma(m+n+a+1) (b-1)^{n+m}} 
        { m! n! b^{m+n+a+1} }
{ }_2F_1 \left( -m, -n; \: -m-n-a; \: \frac{ b(b-2) }{ (b-1)^2 } 
	\right),
\end{align}
where ${}_2F_1\left( \right)$ is the hypergeometric function, and hence, 
\begin{align}\label{app:eigvalue_perturb}
   W_{\fouriersum,\eigLindex}
 &= 
   \pi \eigdecay^{\fouriersum-\eigLindex+1} \ktwo{\layer{B}}{\layer{C}} 
   \eignorm{\fouriersum}{\eigLindex}^2
   \frac{ \Gamma\left(\fouriersum + \eigLindex + 1\right)
          \left( \alpha -1 \right)^{2\eigLindex} }
		{ \eigLindex!^2 \alpha^{\fouriersum + \eigLindex + 1} }
    { }_2F_1 \left( -\eigLindex, \eigLindex;\: \fouriersum - \eigLindex;\: 
		              \frac{ \alpha (\alpha-2) }{ (\alpha-1)^2 } 
		     \right) . 
\end{align}
Since only the diagonal terms are non-zero, the shape of the perturbed eigenfunctions 
remains the same as those for the simplified learning equation, given in 
\eq{\ref{eq:angularWeightFunction}}, but the eigenvalues change according to 
\begin{align}\label{app:lambda_k2}
    \eigvalueFull{\fouriersum}{\eigLindex} 
  &=
    \eigvalueSimple{\fouriersum}{\eigLindex} + W_{\fouriersum,\eigLindex}.
\end{align}

   \section{Derivation of Cartesian eigenfunctions}\label{app:CartesianEigen}
	   
To characterise learning in terms of the eigenfunctions it is useful to approximate 
the system in \eq{\ref{eq:learningRule_Linsker}} by its continuous limit, and initially 
simplify the system by assuming that $\ktwo{}{}=0$. In this case we need to solve 
the eigenvalue problem that integrates the expected covariance over the layer, weighted 
by the probability of connection to the postsynaptic neuron, \postN{}, in layer 
\layer{C}. That is, we need to solve the following eigenfunction equation, 
\begin{align}\label{eq:Linsker_eigfnLearningEqn}
    \eigvalue{}{} \etalayer{} \Lwgtcont{\poscont}{}
 =& \,  
    \cxcoeff 
    \int_{-\infty}^{\infty} \int_{-\infty}^{\infty} 
       \lcov{\Lrate{\postNtwo}{\layer{B}}{} }{ \Lrate{\postN}{\layer{B}}{} }
       \exp{- \frac{ \abs{ \postwocont}^2 } { \radiusvar{\layer{B}}{\layer{C}} }}
       \exp{- \frac{ \abs{ \poscont}^2 } { \radiusvar{\layer{B}}{\layer{C}} }}
       \Lwgtcont{\postwocont}{}
	   d^2 \postwocont												\notag \\
 =&  \, 
    \cxcoeff 
    \int_{-\infty}^{\infty} \int_{-\infty}^{\infty} 
       \exp{- \frac{\abs{\poscont - \postwocont}^2} 
		           {2 \radiusvar{\inputlayer}{\layer{B}} } } 
       \exp{- \frac{ \abs{ \postwocont}^2 } { \radiusvar{\layer{B}}{\layer{C}} }}
       \exp{- \frac{ \abs{ \poscont}^2 } { \radiusvar{\layer{B}}{\layer{C}} }}
       \Lwgtcont{\postwocont}{}
	   d^2 \postwocont												\notag \\
 =&  \, 
    \cxcoeff \, 
    \exp{- \left(\posxcont^2 + \posycont^2 \right)
		   \frac{2\radiusvar{\inputlayer}{\layer{B}} + \radiusvar{\layer{B}}{\layer{C}}}
		        {2 \radiusvar{\inputlayer}{\layer{B}}  \radiusvar{\layer{B}}{\layer{C}}}} \notag \\
 &  \phantom{\cxcoeff} 
    \int_{-\infty}^{\infty} 
    \int_{-\infty}^{\infty} 
	    \exp{- \frac{(2\radiusvar{\inputlayer}{\layer{B}} 
				     + \radiusvar{\layer{B}}{\layer{C}}) \posxtwocont^2
				    + 2\radiusvar{\layer{B}}{\layer{C}} \posxcont \posxtwocont }
				    {2\radiusvar{\inputlayer}{\layer{B}} \radiusvar{\layer{B}}{\layer{C}}}} 
	    \exp{- \frac{(2\radiusvar{\inputlayer}{\layer{B}} 
			         + \radiusvar{\layer{B}}{\layer{C}}) \posytwocont^2
				    + 2\radiusvar{\layer{B}}{\layer{C}} \posycont \posytwocont }
				    {2\radiusvar{\inputlayer}{\layer{B}} \radiusvar{\layer{B}}{\layer{C}}}} 
        \Lwgtcont{\postwocont}{}
	    d\posxtwocont  d\posytwocont \, ,
\end{align}
where $\Lwgtcont{}{}$ is the continuous time approximation to $\Lweight{}{}{}{}$, 
and neuron \preN{} in layer \layer{B} is denoted by its position vector 
$\poscont=\icol{\posxcont}{\posycont}$, $\postwocont = \icol{\posxtwocont}{\posytwocont}$,  
and subscripts have been omitted for readability. The coefficient, \cxcoeff, 
contains the constant terms from the synaptic connection probability 
(\eq{\ref{eq:2DGauss}}), such that  
\begin{align}\label{eq:Linsker_eigEqnCoeff}
    \cxcoeff 
 &= 
	\left( \frac{1}{\pi\radiusvar{\inputlayer}{\layer{B}}} 
	\right)^2. 
\end{align}

Given the separability of the $x$ and $y$ dimensions, in conjunction with the 
exponential weight function, consider the Hermite polynomial as the form of the 
eigenfunction, such that the eigenfunctions are given by 
\begin{align}\label{eq:Linske_eigenvec}
    \eigvec{\eigxindex,\eigyindex}
         {\frac{\posxcont}{\sqrt{\eigdecay}}, \frac{\posycont}{\sqrt{\eigdecay}}}
 &= 
    \eignorm{\eigxindex}{\eigyindex}
    \hermite{\eigxindex}{\frac{\posxcont}{\sqrt{\eigdecay}}} 
    \hermite{\eigyindex}{\frac{\posycont}{\sqrt{\eigdecay}}} 
	\exp{-\frac{\posxcont^2 + \posycont^2}{2\eigdecay}},
\end{align}
where $\eigxindex$ and $\eigyindex$ denote the order of the polynomial for each 
dimension, giving a two-dimensional eigenfunction of order  
$\eigxindex + \eigyindex$, and  
${\eignorm{\eigxindex}{\eigyindex} = \hermitecoeff{\eigxindex} \hermitecoeff{\eigyindex}}$ 
is a normalization constant \citep{Rom84}. $\eigdecay$ is a parameter that 
must be determined. 

Consequently, when this expression is input in to the eigenfunction equation from 
\eq{\ref{eq:Linsker_eigfnLearningEqn}} the learning equation becomes, 
\begin{align}\label{eq:radialCartesian_eig_k2=0}
    \etalayer{} \eigvalue{\eigxindex}{\eigyindex}
	\eigvec{\eigxindex,\eigyindex} 
         {\frac{\posxcont}{\sqrt{\eigdecay}}, \frac{\posycont}{\sqrt{\eigdecay}}}
 =& 
    \cxcoeff \: \exp{- \left(\posxcont^2 + \posycont^2 \right)
		               \frac{2\radiusvar{\inputlayer}{\layer{B}} 
					        + \radiusvar{\layer{B}}{\layer{C}}}
		                   {2 \radiusvar{\inputlayer}{\layer{B}}  
					  	      \radiusvar{\layer{B}}{\layer{C}}}}  
    \int\displaylimits_{-\infty}^{\infty} 
    \int\displaylimits_{-\infty}^{\infty} 
        \exp{-\frac{(2\radiusvar{\inputlayer}{\layer{B}} 
				    + \radiusvar{\layer{B}}{\layer{C}}) \posxtwocont^2
				    +2\radiusvar{\layer{B}}{\layer{C}}  \posxcont \posxtwocont }
				   { 2\radiusvar{\inputlayer}{\layer{B}} 
					  \radiusvar{\layer{B}}{\layer{C}}} } \notag \\
	&   \exp{-\frac{(2\radiusvar{\inputlayer}{\layer{B}} 
				    + \radiusvar{\layer{B}}{\layer{C}}) \posytwocont^2
				    +2\radiusvar{\layer{B}}{\layer{C}}  \posycont \posytwocont }
				   { 2\radiusvar{\inputlayer}{\layer{B}} 
				 	  \radiusvar{\layer{B}}{\layer{C}}} } 
		\eigvec{\eigxindex,\eigyindex} 
			 {\frac{\posxtwocont}{\sqrt{\eigdecay}}, 
			  \frac{\posytwocont}{\sqrt{\eigdecay}}}     
	    d\posxtwocont  d\posytwocont \: , 
\end{align} 
which holds true only when 
\begin{align}\label{eq:eigendecay_Cartesian}
    \eigdecay 
 &= \frac{ \radiusvar{\layer{B}}{\layer{C}} }
	     {2\sqrt{1 + \frac{ \radiusvar{\layer{B}}{\layer{C}} }
					      {\radiusvar{\inputlayer}{\layer{B}} } } }.
\end{align}	
Note that this result agrees with the result derived for the radial eigenfunctions, 
\eq{\ref{eq:eigendecay_radial}}, once the scaling of the $\posr$ by 
$\nicefrac{1}{\radius{\inputlayer}{\layer{B}}}$ is accounted for.

Due to the separability of the dimensions, the eigenvalues for $\posxcont$ and 
$\posycont$ can be derived independently. Therefore initially consider the 
problem in just one dimension. In deriving the eigenvalues for the complete 
orthogonal set of Hermite polynomials, we follow the procedure used in 
\citep{WimGerHem98}, and make the Ansatz that, 
\begin{align}\label{eq:radialEigenval}
   \eigvalue{\eigxindex}{} = \Lambda q^{\eigxindex}.
\end{align}

Using the generating function for one dimensional Hermite polynomials, 
\begin{align}\label{eq:HermiteGenFn}
   \exp{-t^2 + 2t\frac{\posx{}{}}{\sqrt{\eigdecay}}}
 &= 
   \Sum{\fouriersum}{0}{\infty} 
      \frac{ t^{\fouriersum} }{\fouriersum !}
	  \hermite{\fouriersum}{\frac{\posx{}{}}{\sqrt{\eigdecay}}},
\end{align}
and the orthogonality of Hermite polynomials with respect to a Gaussian 
weight function, we know that, 
\begin{align}\label{eq:HermiteGenWithWeightFn}
   \exp{-t^2 + 2t\frac{\posx{}{}}{\sqrt{\eigdecay}}} 
   \exp{-\frac{\posx{}{}^2}{2\eigdecay}} 
 &= 
   \Sum{\fouriersum}{0}{\infty} 
      \frac{ t^{\fouriersum} }{\fouriersum !}
	  \hermite{\fouriersum}{\frac{\posx{}{}}{\sqrt{\eigdecay}}}
      \exp{-\frac{\posx{}{}^2}{2\eigdecay}}.
\end{align}
Combining this with \eq{\ref{eq:radialCartesian_eig_k2=0}}, considering a single 
dimension only, gives, 
\begin{align}\label{eq:Cartesian_generatingFn}
   \Sum{\fouriersum}{0}{\infty} \eigvalue{\fouriersum}{}
      \frac{ t^{\fouriersum} }{\fouriersum !}
	  \hermite{\fouriersum}{\frac{\posx{}{}}{\sqrt{\eigdecay}}}
      \exp{-\frac{\posx{}{}^2}{2\eigdecay}} 
 =\, & 
    \sqrt{\cxcoeff} \exp{- \posxcont^2 
		            \left( \frac{2\radiusvar{\inputlayer}{\layer{B}} 
							     + \radiusvar{\layer{B}}{\layer{C}}}
		                        {2 \radiusvar{\inputlayer}{\layer{B}}  
								   \radiusvar{\layer{B}}{\layer{C}}} 
					\right) }                                                \notag \\
 &  \int_{-\infty}^{\infty} 
       \exp{- \frac{ \left( \left(2\radiusvar{\inputlayer}{\layer{B}} 
				                 + \radiusvar{\layer{B}}{\layer{C}} \right) 
			                \posintx^2 
				         + 2\radiusvar{\layer{B}}{\layer{C}} \posintx{}{} \posx{}{} 
			         \right) }
			       {2 \radiusvar{\inputlayer}{\layer{B}} \radiusvar{\layer{B}}{\layer{C}}} }
	   \Sum{\fouriersum}{0}{\infty} 
		  \frac{ t^{\fouriersum} }{\fouriersum !}
		  \hermite{\fouriersum}{\frac{\posintx}{\sqrt{\eigdecay}}}
		  \exp{-\frac{\posx{}{}^2}{2\eigdecay}}
	  d\posintx																\notag \\
 =\, & 
    \sqrt{\cxcoeff} \exp{- \posxcont^2 
		            \left( \frac{2\radiusvar{\inputlayer}{\layer{B}} 
							    + \radiusvar{\layer{B}}{\layer{C}}}
		                        {2\radiusvar{\inputlayer}{\layer{B}}  
								  \radiusvar{\layer{B}}{\layer{C}}} 
					\right) }												\notag \\
 &  \int_{-\infty}^{\infty} 
       \exp{- \frac{ \left( \left(2\radiusvar{\inputlayer}{\layer{B}} 
				                 + \radiusvar{\layer{B}}{\layer{C}} \right) 
			                \posintx^2 
				         + 2\radiusvar{\layer{B}}{\layer{C}} \posintx{}{} \posx{}{} 
			         \right) }
			       {2 \radiusvar{\inputlayer}{\layer{B}} \radiusvar{\layer{B}}{\layer{C}}} }
	   \exp{ -t^2 + 2t\frac{\posintx}{\sqrt{\eigdecay}} } 
	  \exp{-\frac{\posx{}{}^2}{2\eigdecay}}
		 d\posintx
\end{align}

Evaluating the right hand side, 
\begin{align}
    \text{RHS} 
 =\, & 
    \sqrt{\cxcoeff} 
	\exp{- \posxcont^2 
		   \left( \frac{2\radiusvar{\inputlayer}{\layer{B}} 
							     + \radiusvar{\layer{B}}{\layer{C}}}
		               {2 \radiusvar{\inputlayer}{\layer{B}}  
						  \radiusvar{\layer{B}}{\layer{C}}}  
		   \right) }  \notag \\
  & \int_{-\infty}^{\infty} 
       \exp{- \frac{ \left( \left( 2\eigdecay \radiusvar{\inputlayer}{\layer{B}} 
				                  + \eigdecay \radiusvar{\layer{B}}{\layer{C}} 
						          + \radiusvar{\inputlayer}{\layer{B}} 
								    \radiusvar{\layer{B}}{\layer{C}}
							\right) \posintx^2 
				         + 2\eigdecay \radiusvar{\layer{B}}{\layer{C}} \posintx{}{} \posx{}{} 
			         \right) }
			       {2 \eigdecay \radiusvar{\inputlayer}{\layer{B}} 
					            \radiusvar{\layer{B}}{\layer{C}} } 
			}
	   \exp{ -t^2 + 2t\frac{\posintx}{\sqrt{\eigdecay}} }
	  d\posintx  \notag \\
 =\, & 
    \sqrt{\cxcoeff} \exp{ -t^2} 
	\exp{- \posxcont^2 
		   \left( \frac{2\radiusvar{\inputlayer}{\layer{B}} + \radiusvar{\layer{B}}{\layer{C}}}
		               {2 \radiusvar{\inputlayer}{\layer{B}}  \radiusvar{\layer{B}}{\layer{C}}} 
		   \right) }  \notag \\
 &  \int_{-\infty}^{\infty} 
       \exp{- \frac{ \left( 2\eigdecay \radiusvar{\inputlayer}{\layer{B}} 
						   + \eigdecay \radiusvar{\layer{B}}{\layer{C}} 
						   + \radiusvar{\inputlayer}{\layer{B}} \radiusvar{\layer{B}}{\layer{C}}
					 \right) \posintx^2 
				  +  \left( 2\eigdecay \radiusvar{\layer{B}}{\layer{C}} \posx{}{} 
						  + 4t \sqrt{\eigdecay} \radiusvar{\inputlayer}{\layer{B}} 
						       \radiusvar{\layer{B}}{\layer{C}}
				     \right) \posintx }
			       {2 \eigdecay \radiusvar{\inputlayer}{\layer{B}} 
					            \radiusvar{\layer{B}}{\layer{C}}} }
	  d\posintx   \notag \\
 =\, & 
    \sqrt{\cxcoeff} \exp{ -t^2} 
	\exp{- \posxcont^2 \left( \frac{2 \radiusvar{\inputlayer}{\layer{B}} 
						 	        + \radiusvar{\layer{B}}{\layer{C}}}
		                           {2 \radiusvar{\inputlayer}{\layer{B}}  
								      \radiusvar{\layer{B}}{\layer{C}}}
					   \right)
		}
    \exp{ \frac{ \left( \eigdecay \radiusvar{\layer{B}}{\layer{C}} \posx{}{} 
				 + 2t \sqrt{\eigdecay} \radiusvar{\inputlayer}{\layer{B}} 
				                       \radiusvar{\layer{B}}{\layer{C}}
		 	     \right)^2 }
			   { \left(  2\eigdecay \radiusvar{\inputlayer}{\layer{B}} 
					  + \eigdecay \radiusvar{\layer{B}}{\layer{C}} 
					  + \radiusvar{\inputlayer}{\layer{B}} 
					    \radiusvar{\layer{B}}{\layer{C}}
			     \right)
			    2\eigdecay \radiusvar{\inputlayer}{\layer{B}} 
			               \radiusvar{\layer{B}}{\layer{C}} } 
		}   \notag \\
 &  \int_{-\infty}^{\infty} 
	   \exp{- \frac{1}{2 \eigdecay \radiusvar{\inputlayer}{\layer{B}} 
		                           \radiusvar{\layer{B}}{\layer{C}}} 
		      \left( \left( 2\eigdecay \radiusvar{\inputlayer}{\layer{B}} 
							  + \eigdecay \radiusvar{\layer{B}}{\layer{C}} 
							  + \radiusvar{\inputlayer}{\layer{B}} 
							    \radiusvar{\layer{B}}{\layer{C}}
					 \right)^{\frac{1}{2}} \posintx				
				   -\frac{ \eigdecay \radiusvar{\layer{B}}{\layer{C}} \posx{}{} 
						 + 2t \sqrt{\eigdecay} \radiusvar{\inputlayer}{\layer{B}} 
						      \radiusvar{\layer{B}}{\layer{C}} }
				         {\left(2\eigdecay \radiusvar{\inputlayer}{\layer{B}} 
							 + \eigdecay \radiusvar{\layer{B}}{\layer{C}} 
							 + \radiusvar{\inputlayer}{\layer{B}} 
							   \radiusvar{\layer{B}}{\layer{C}}
					       \right)^{\frac{1}{2})} }
			  \right)^2 
			}
	  d\posintx   \notag \\
 =\, & 
    \sqrt{\cxcoeff} \exp{ -t^2} 
	\left( \frac{2 \pi \eigdecay \radiusvar{\inputlayer}{\layer{B}} 
			                     \radiusvar{\layer{B}}{\layer{C}}}
                {2 \eigdecay \radiusvar{\inputlayer}{\layer{B}} 
				 + \eigdecay \radiusvar{\layer{B}}{\layer{C}} 
				 + \radiusvar{\inputlayer}{\layer{B}} \radiusvar{\layer{B}}{\layer{C}} }
  	\right) \notag \\
 &	\exp{- \posxcont^2 \left( \frac{2 \radiusvar{\inputlayer}{\layer{B}} 
			                        + \radiusvar{\layer{B}}{\layer{C}}}
		                           {2 \radiusvar{\inputlayer}{\layer{B}}  
								      \radiusvar{\layer{B}}{\layer{C}}}
					   \right) 
		}
    \exp{ \frac{ \left( \eigdecay \radiusvar{\layer{B}}{\layer{C}} \posx{}{} 
				   + 2t \sqrt{\eigdecay} \radiusvar{\inputlayer}{\layer{B}} 
				                       \radiusvar{\layer{B}}{\layer{C}}
			     \right)^2 }
			   { \left(  2\eigdecay \radiusvar{\inputlayer}{\layer{B}} 
					    + \eigdecay \radiusvar{\layer{B}}{\layer{C}} 
					    + \radiusvar{\inputlayer}{\layer{B}} \radiusvar{\layer{B}}{\layer{C}}
			     \right)
			    2\eigdecay \radiusvar{\inputlayer}{\layer{B}} 
			               \radiusvar{\layer{B}}{\layer{C}} } 
		} \, .
\end{align}

Evaluating the left hand side of \eq{\ref{eq:Cartesian_generatingFn}} gives 
\begin{align}
   \text{LHS} 
 &=\,  
   \Sum{\fouriersum}{0}{\infty} \eigAllCoeff 
      \frac{ (qt)^{\fouriersum} }{\fouriersum !}
	  \hermite{\fouriersum}{\frac{\posx{}{}}{\sqrt{\eigdecay}}}
      \exp{-\frac{\posx{}{}^2}{2\eigdecay}}									\notag \\
 &=\,  
   \eigAllCoeff \exp{-(qt)^2 + \frac{2q}{\sqrt{\eigdecay}}\posx{}{}t} 
   \exp{-\frac{\posx{}{}^2}{2\eigdecay}}. 
\end{align}
Comparing RHS and LHS requires that 
\begin{align}\label{eq:q_genFn}
	q =\,  \frac{\eigdecay \radiusvar{\layer{B}}{\layer{C}}}
             {2\radiusvar{\inputlayer}{\layer{B}} \eigdecay + \radiusvar{\layer{B}}{\layer{C}} \eigdecay 
			 + \radiusvar{\inputlayer}{\layer{B}} \radiusvar{\layer{B}}{\layer{C}}},
\end{align}
and 
\begin{align}\label{eq:eigenvalue_coeff}
    \eigAllCoeff 
  &=\,  
    \sqrt{ \frac{ 2\pi \eigdecay \radiusvar{\inputlayer}{\layer{B}} 
	                             \radiusvar{\layer{B}}{\layer{C}} }
				{ 2 \eigdecay \radiusvar{\inputlayer}{\layer{B}} 
				+ 2 \eigdecay \radiusvar{\layer{B}}{\layer{C}} 
				+ \radiusvar{\inputlayer}{\layer{B}} 
			  	  \radiusvar{\layer{B}}{\layer{C}} } }  \notag \\
  &=\, 
	\sqrt{ 2\pi \radiusvar{\inputlayer}{\layer{B}} q}
\end{align}
so that 
\begin{align}\label{eq:eigenvalue_1D}
   \eigvalue{\fouriersum}{}
 &= 
   \eigAllCoeff q^{\fouriersum}  \notag \\
 &= 
   \sqrt{ 2\pi \radiusvar{\inputlayer}{\layer{B}}} q^{\fouriersum + \frac{1}{2}}.
\end{align}

For two dimensions, the final eigenfunctions and eigenvalue pairs, for order 
$\eigxindex$ and $\eigyindex$, for the $x$ and $y$ dimensions respectively, 
are given by, 
\begin{subequations}\label{app:eig_k2=0}
\begin{align}
   \eigvalue{\eigxindex}{\eigyindex}
 &= 
   2\pi\radiusvar{\inputlayer}{\layer{B}} q^{\eigxindex + \eigyindex + 1} 
   \label{app:cartesianEigenval_simple} \\
    \eigvec{\eigxindex,\eigyindex}
         {\frac{\posxcont}{\sqrt{\eigdecay}}, \frac{\posycont}{\sqrt{\eigdecay}}}
 &= 
    \hermitecoeff{\eigxindex} \hermitecoeff{\eigyindex} 
    \hermite{\eigxindex}{\frac{\posxcont}{\sqrt{\eigdecay}}} 
    \hermite{\eigyindex}{\frac{\posycont}{\sqrt{\eigdecay}}} 
	\exp{-\frac{\posxcont^2 + \posycont^2}{2\eigdecay}} \, . 
   \label{app:cartesianEigenfn_simple} 
\end{align}
\end{subequations}

   \section{Expected number of shared inputs with radially dependent cell density}
   \label{app:sharedInput_radial}
	   We consider the expected number of shared inputs between two cells in 
layer $\layer{B}$ when there is radially dependent cell density, such 
that cell density is highest in the centre of the layer, and decreases 
linearly with distance from the centre. In this case, an assumed 
consequence is a reduction in the radius of the synaptic connectivity 
distribution in the layer centre, with a corresponding linear increase 
with radial distance traversed from the layer centre. 

The probability of presynaptic neuron, $\preN$, in layer 
$\layer{\inputlayer}$, generating a synaptic connection to postsynaptic neuron, 
$\postN$, in layer $\layer{B}$, is given in \eq{\ref{eq:2DGauss}}. 

If neuron \postNtwo{} is \distance{\postNtwo}{\centreN}{\layer{B}}{\layer{B}} far 
from the layer centre, with a cell connection density of \radius{\postN}{}, 
then the expression for the expected number of neurons in layer \layer{\inputlayer} 
connecting to both neuron \postN{} and neuron \postNtwo{}, in the continuous 
limit, is given by 
\begin{align}\label{eq:joint_radial_density}
     \expect{ \Nshared{\layer{A}}{\layer{B}} ; 
  	          \posvec{\posx{\postN}{}}{\posy{\postNtwo}{}},
	          \posvec{\posx{\postN}{}}{\posy{\postNtwo}{}} }
 &=
     \frac{ (\N{\layer{A}}{\layer{B}})^2 }
		  { \pi^2 \left( \radius{\layer{A}}{\layer{B}} \right)^4
	        \distance{\postN}{}{}{}^2 \distance{\postNtwo}{}{}{}^2 } 
   { \mathlarger{\iint\limits_{\posx{}{}\posy{}{}}} }
	 \exp{- \frac{ \left( \posx{}{} - \posx{\postN}{} \right)^2 + 
		           \left( \posy{}{} - \posy{\postN}{} \right)^2}
				 { \distance{\postN}{}{}{}^2 \radiusvar{\layer{A}}{\layer{B}} } }
	 \exp{- \frac{ \left( \posx{}{} - \posx{\postNtwo}{} \right)^2 + 
		           \left( \posy{}{} - \posy{\postNtwo}{} \right)^2}
				 { \distance{\postNtwo}{}{}{}^2 \radiusvar{\layer{A}}{\layer{B}} } }
   \,d\posx{}{} \,d\posy{}{},                                     \notag \\
  &=
    \frac{ (\N{\layer{A}}{\layer{B}})^2 }
		 { \pi^2 \left( \radius{\layer{A}}{\layer{B}} \right)^4
	      \distance{\postN}{}{}{}^2 \distance{\postNtwo}{}{}{}^2 }
  { \mathlarger{\int\limits_{\posx{}{}}} }
	\exp{- \frac{ \distance{\postNtwo}{}{}{}^2
		           \left( \posx{}{} - \posx{\postN}{} \right)^2 + 
                   \distance{\postN}{}{}{}^2
		           \left( \posx{}{} - \posx{\postNtwo}{} \right)^2}
			     { \distance{\postNtwo}{}{}{}^2 
				   \distance{\postN}{}{}{}^2 
				   \radiusvar{\layer{A}}{\layer{B}}} }
  \,d\posx{}{} 
   { \mathlarger{\int\limits_{\posy{}{}}} }
	 \exp{- \frac{ \distance{\postNtwo}{}{}{}^2 
		           \left( \posy{}{} - \posy{\postN}{} \right)^2 + 
                   \distance{\postN}{}{}{}^2
		           \left( \posy{}{} - \posy{\postNtwo}{} \right)^2 }
			     { \distance{\postN}{}{}{}^2 
                   \distance{\postNtwo}{}{}{}^2
				   \radiusvar{\layer{A}}{\layer{B}}} }
   \,d\posy{}{},
\end{align}
where we drop the centre subscript, \centreN{}, and layer superscript, 
$^{\layer{B}\layer{B}}$,
for readability. The exponent in the left sum can be simplified as 
\begin{align*}
  \lefteqn{ \frac{ \distance{\postNtwo}{}{}{}^2
         		   \left( \posx{}{} - \posx{\postN}{} \right)^2 + 
                   \distance{\postN}{}{}{}^2
        		   \left( \posx{}{} - \posx{\postNtwo}{} \right)^2}
        	 	 { \distance{\postNtwo}{}{}{}^2 
        		   \distance{\postN}{}{}{}^2 
        		   \radiusvar{\layer{A}}{\layer{B}} }  }   \notag \\             
 & \verylargespace \verylargespace =
   \frac{ \left( \distance{\postN}{}{}{}^2 
		       + \distance{\postNtwo}{}{}{}^2 \right) \posx{}{}^2 
        -2\left( \distance{\postNtwo}{}{}{}^2 \posx{\postN}{} 
			   + \distance{\postN}{}{}{}^2 \posx{\postNtwo}{} \right) \posx{}{} 
        + \left( \distance{\postNtwo}{}{}{}^2 \posx{\postN}{}^2 
			   + \distance{\postN}{}{}{}^2 \posx{\postNtwo}{}^2 \right) }
        {\distance{\postN}{}{}{}^2 \distance{\postNtwo}{}{}{}^2
		 \radiusvar{\layer{A}}{\layer{B}} }                 \notag \\
 & \verylargespace \verylargespace = 
   \frac{\distance{\postN}{}{}{}^2 + \distance{\postNtwo}{}{}{}^2}
        {\distance{\postN}{}{}{}^2 \distance{\postNtwo}{}{}{}^2
		 \radiusvar{\layer{A}}{\layer{B}} }
   \left[\posx{}{}^2 
       - \frac{2\left( \distance{\postNtwo}{}{}{}^2 \posx{\postN}{} 
			         + \distance{\postN}{}{}{}^2 \posx{\postNtwo}{} 
				\right) }
              {\distance{\postN}{}{}{}^2 + \distance{\postNtwo}{}{}{}^2}  
		 \posx{}{}
	   + \frac{\distance{\postNtwo}{}{}{}^2 \posx{\postN}{}^2 
			 + \distance{\postN}{}{}{}^2 \posx{\postNtwo}{}^2}
              {\distance{\postN}{}{}{}^2 + \distance{\postNtwo}{}{}{}^2}
   \right]												    \notag \\
 & \verylargespace \verylargespace = 
   \frac{\distance{\postN}{}{}{}^2 + \distance{\postNtwo}{}{}{}^2}
        {\distance{\postN}{}{}{}^2 \distance{\postNtwo}{}{}{}^2
		 \radiusvar{\layer{A}}{\layer{B}} }
   \left[\left( \posx{}{}
              - \frac{\distance{\postNtwo}{}{}{}^2 \posx{\postN}{} 
	 		               + \distance{\postN}{}{}{}^2 \posx{\postNtwo}{} }
                     {\distance{\postN}{}{}{}^2 + \distance{\postNtwo}{}{}{}^2}
		 \right)^2
       - \left(\frac{ \distance{\postNtwo}{}{}{}^2 \posx{\postN}{} 
	 		        + \distance{\postN}{}{}{}^2 \posx{\postNtwo}{} }
                    {\distance{\postN}{}{}{}^2 + \distance{\postNtwo}{}{}{}^2}
		 \right)^2
	   + \frac{\distance{\postNtwo}{}{}{}^2 \posx{\postN}{}^2 
			 + \distance{\postN}{}{}{}^2 \posx{\postNtwo}{}^2}
              {\distance{\postN}{}{}{}^2 + \distance{\postNtwo}{}{}{}^2}
   \right]														\notag \\
 & \verylargespace \verylargespace = 
   \frac{\distance{\postN}{}{}{}^2 + \distance{\postNtwo}{}{}{}^2}
        {\distance{\postN}{}{}{}^2 \distance{\postNtwo}{}{}{}^2
		 \radiusvar{\layer{A}}{\layer{B}} }
   \left[\left( \posx{}{}
              - \frac{\distance{\postNtwo}{}{}{}^2 \posx{\postN}{} 
	 		               + \distance{\postN}{}{}{}^2 \posx{\postNtwo}{} }
                     {\distance{\postN}{}{}{}^2 + \distance{\postNtwo}{}{}{}^2}
		 \right)^2
       - \frac{\distance{\postNtwo}{}{}{}^4 \posx{\postN}{}^2 
	 		 + \distance{\postN}{}{}{}^4 \posx{\postNtwo}{}^2 
	         +2\distance{\postN}{}{}{}^2 \distance{\postNtwo}{}{}{}^2 
	           \posx{\postN}{} \posx{\postNtwo}{}}
              {\left(\distance{\postN}{}{}{}^2 + \distance{\postNtwo}{}{}{}^2\right)^2}
	   + \frac{\distance{\postNtwo}{}{}{}^2 \posx{\postN}{}^2 
			 + \distance{\postN}{}{}{}^2 \posx{\postNtwo}{}^2}
              {\distance{\postN}{}{}{}^2 + \distance{\postNtwo}{}{}{}^2}
   \right]														\notag \\
 & \verylargespace \verylargespace = 
   \frac{\distance{\postN}{}{}{}^2 + \distance{\postNtwo}{}{}{}^2}
        {\distance{\postN}{}{}{}^2 \distance{\postNtwo}{}{}{}^2
		 \radiusvar{\layer{A}}{\layer{B}} }
   \left[\left( \posx{}{}
              - \frac{\distance{\postNtwo}{}{}{}^2 \posx{\postN}{} 
	 		               + \distance{\postN}{}{}{}^2 \posx{\postNtwo}{} }
                     {\distance{\postN}{}{}{}^2 + \distance{\postNtwo}{}{}{}^2}
		 \right)^2
       - \frac{\left(\distance{\postNtwo}{}{}{}^2 + \distance{\postNtwo}{}{}{}^2\right)^2
	 		   \left(\distance{\postNtwo}{}{}{}^2 \posx{\postN}{}^2 
			       + \distance{\postN}{}{}{}^2 \posx{\postNtwo}{}^2 \right)
	         +2\distance{\postN}{}{}{}^2 \distance{\postNtwo}{}{}{}^2
	           \posx{\postN}{} \posx{\postNtwo}{} 
	         - \distance{\postN}{}{}{}^2 \distance{\postNtwo}{}{}{}^2\posx{\postN}{}
	         - \distance{\postN}{}{}{}^2 \distance{\postNtwo}{}{}{}^2\posx{\postNtwo}{} }
              {\left(\distance{\postN}{}{}{}^2 + \distance{\postNtwo}{}{}{}^2\right)^2}
	   + \frac{\distance{\postNtwo}{}{}{}^2 \posx{\postN}{}^2 
			 + \distance{\postN}{}{}{}^2 \posx{\postNtwo}{}^2}
              {\distance{\postN}{}{}{}^2 + \distance{\postNtwo}{}{}{}^2}
   \right]														\notag \\
 & \verylargespace \verylargespace = 
   \frac{\distance{\postN}{}{}{}^2 + \distance{\postNtwo}{}{}{}^2}
        {\distance{\postN}{}{}{}^2 \distance{\postNtwo}{}{}{}^2
		 \radiusvar{\layer{A}}{\layer{B}} }
   \left[\left( \posx{}{}
              - \frac{\distance{\postNtwo}{}{}{}^2 \posx{\postN}{} 
	 		        + \distance{\postN}{}{}{}^2 \posx{\postNtwo}{} }
                     {\distance{\postN}{}{}{}^2 + \distance{\postNtwo}{}{}{}^2}
		 \right)^2
       - \frac{\distance{\postNtwo}{}{}{}^2 \posx{\postN}{}^2 
			 + \distance{\postN}{}{}{}^2 \posx{\postNtwo}{}^2 }
              {\distance{\postN}{}{}{}^2 + \distance{\postNtwo}{}{}{}^2 }
	   + \frac{\distance{\postNtwo}{}{}{}^2 \posx{\postN}{}^2 
			 + \distance{\postN}{}{}{}^2 \posx{\postNtwo}{}^2}
              {\distance{\postN}{}{}{}^2 + \distance{\postNtwo}{}{}{}^2}
	   - \frac{\distance{\postN}{}{}{}^2 \distance{\postNtwo}{}{}{}^2 
	           \left( \posx{\postN}{}^2 + \posx{\postNtwo}{}^2
	                -2\posx{\postN}{} \posx{\postNtwo}{} \right)}
              {\left( \distance{\postN}{}{}{}^2 + \distance{\postNtwo}{}{}{}^2 \right)^2}
   \right]														\notag \\
 & \verylargespace \verylargespace =
   \frac{\distance{\postN}{}{}{}^2 + \distance{\postNtwo}{}{}{}^2}
        {\distance{\postN}{}{}{}^2 \distance{\postNtwo}{}{}{}^2
		 \radiusvar{\layer{A}}{\layer{B}} }
   \left( \posx{}{}
        - \frac{\distance{\postNtwo}{}{}{}^2 \posx{\postN}{} 
	 		  + \distance{\postN}{}{}{}^2 \posx{\postNtwo}{} }
               {\distance{\postN}{}{}{}^2 + \distance{\postNtwo}{}{}{}^2}
   \right)^2
 - \left( \frac{\distance{\postN}{}{}{}^2 + \distance{\postNtwo}{}{}{}^2}
               {\distance{\postN}{}{}{}^2 \distance{\postNtwo}{}{}{}^2
		 \radiusvar{\layer{A}}{\layer{B}} } \right)
   \frac{\distance{\postN}{}{}{}^2 \distance{\postNtwo}{}{}{}^2 
	     \left( \posx{\postN}{}^2 + \posx{\postNtwo}{}^2
	          -2\posx{\postN}{} \posx{\postNtwo}{} \right)}
        {\left( \distance{\postN}{}{}{}^2 + \distance{\postNtwo}{}{}{}^2 \right)^2} 
																			\notag \\
 & \verylargespace \verylargespace =
   \frac{\distance{\postN}{}{}{}^2 + \distance{\postNtwo}{}{}{}^2}
        {\distance{\postN}{}{}{}^2 \distance{\postNtwo}{}{}{}^2
		 \radiusvar{\layer{A}}{\layer{B}} }
   \left( \posx{}{}
              - \frac{\distance{\postNtwo}{}{}{}^2 \posx{\postN}{} 
	 		               + \distance{\postN}{}{}{}^2 \posx{\postNtwo}{} }
                     {\distance{\postN}{}{}{}^2 + \distance{\postNtwo}{}{}{}^2}
   \right)^2
 - \frac{\left( \posx{\postN}{} - \posx{\postNtwo}{} \right)^2 }
		{\radiusvar{\layer{A}}{\layer{B}}
	   	 \left( \distance{\postN}{}{}{}^2 + \distance{\postNtwo}{}{}{}^2 \right)}.
\end{align*}
Applying this result to the $\posy{}{}$ integrand in \eq{\ref{eq:joint_radial_density}}
also, we get
\begin{align}\label{app:radialSharedConns}
\begin{split}
   \expect{\Nshared{\layer{A}}{\layer{B}}; 
	       \posvec{\posx{\postN}{}}{\posy{\postN}{}},
	       \posvec{\posx{\postN}{}}{\posy{\postN}{}} }
  =&
	\frac{ (\N{\layer{A}}{\layer{B}})^2}
		 { \pi^2 \left( \radius{\layer{A}}{\layer{B}} \right)^4
	       \distance{\postN}{}{}{}^2 \distance{\postNtwo}{}{}{}^2 }
   {\mathlarger{\int\limits_{\posx{}{}}}}
     \exp{- \frac{\distance{\postN}{}{}{}^2 + \distance{\postNtwo}{}{}{}^2}
                 {\distance{\postN}{}{}{}^2 \distance{\postNtwo}{}{}{}^2
				  \radiusvar{\layer{A}}{\layer{B}} }
            \left( \posx{}{}
                 - \frac{\distance{\postNtwo}{}{}{}^2 \posx{\postN}{} 
	 		           + \distance{\postN}{}{}{}^2 \posx{\postNtwo}{} }
                        {\distance{\postN}{}{}{}^2 + \distance{\postNtwo}{}{}{}^2}
            \right)^2
          - \frac{\left( \posx{\postN}{} - \posx{\postNtwo}{} \right)^2 }
				 {\radiusvar{\layer{A}}{\layer{B}}
 	   	          \left( \distance{\postN}{}{}{}^2 
					   + \distance{\postNtwo}{}{}{}^2 \right)}}
  \,d\posx{}{}															\\
  &\phantom{\frac{(\N{\layer{A}}{\layer{B}})^2}
				 {\pi^2 \left( \radius{\layer{A}}{\layer{B}} \right)^4
	              \distance{\postN}{}{}{}^2 \distance{\postNtwo}{}{}{}^2} }
  {\mathlarger{\int\limits_{\posy{}{}}}}
     \exp{- \frac{\distance{\postN}{}{}{}^2 + \distance{\postNtwo}{}{}{}^2}
                 {\distance{\postN}{}{}{}^2 \distance{\postNtwo}{}{}{}^2
				  \radiusvar{\layer{A}}{\layer{B}} }
            \left( \posy{}{}
                 - \frac{\distance{\postNtwo}{}{}{}^2 \posy{\postN}{} 
	 		           + \distance{\postN}{}{}{}^2 \posy{\postNtwo}{} }
                        {\distance{\postN}{}{}{}^2 + \distance{\postNtwo}{}{}{}^2}
            \right)^2
          - \frac{\left( \posy{\postN}{} - \posy{\postNtwo}{} \right)^2 }
				 {\radiusvar{\layer{A}}{\layer{B}}
 	   	          \left( \distance{\postN}{}{}{}^2 
					   + \distance{\postNtwo}{}{}{}^2 \right)}}
   \,d\posy{}{}
\end{split}																\notag \\
 =&
	\frac{ (\N{\layer{A}}{\layer{B}})^2 }
		 { \pi \left( \radius{\layer{A}}{\layer{B}} \right)^4
	       \distance{\postN}{}{}{}^2 \distance{\postNtwo}{}{}{}^2 }
   \exp{- \frac{\left( \posx{\postN}{} - \posx{\postNtwo}{} \right)^2
              + \left( \posy{\postN}{} - \posy{\postNtwo}{} \right)^2 }
			   {\radiusvar{\layer{A}}{\layer{B}}
			    \left(\distance{\postN}{}{}{}^2 
				    + \distance{\postNtwo}{}{}{}^2\right)}}
	\left(\frac{\pi \radiusvar{\layer{A}}{\layer{B}}
		       \distance{\postN}{}{}{}^2 \distance{\postNtwo}{}{}{}^2}
              {\distance{\postN}{}{}{}^2 + \distance{\postNtwo}{}{}{}^2}
   \right)^{\nicefrac{1}{2}}
	\left(\frac{\pi \radiusvar{\layer{A}}{\layer{B}}
		       \distance{\postN}{}{}{}^2 \distance{\postNtwo}{}{}{}^2}
              {\distance{\postN}{}{}{}^2 + \distance{\postNtwo}{}{}{}^2}
   \right)^{\nicefrac{1}{2}}											\notag \\
  =&
   \frac{(\N{\layer{A}}{\layer{B}})^2}
		{\pi \radiusvar{\layer{A}}{\layer{B}}
         \left(\distance{\postN}{}{}{}^2 + \distance{\postNtwo}{}{}{}^2 \right)}
   \exp{-\frac{\left( \posx{\postN}{} - \posx{\postNtwo}{} \right)^2
	         + \left( \posy{\postN}{} - \posy{\postNtwo}{} \right)^2 }
			  {\radiusvar{\layer{A}}{\layer{B}}
			   \left(\distance{\postN}{}{}{}^2 
			       + \distance{\postNtwo}{}{}{}^2\right)}}			\notag \\
  =&
   \frac{(\N{\layer{A}}{\layer{B}})^2}
		{\pi \radiusvar{\layer{A}}{\layer{B}}
         \left(\distance{\postN}{}{}{}^2 + \distance{\postNtwo}{}{}{}^2 \right)}
   \exp{-\frac{\distance{\postN}{\postNtwo}{}{}^2}
			  {\radiusvar{\layer{A}}{\layer{B}}
			   \left(\distance{\postN}{}{}{}^2 
			       + \distance{\postNtwo}{}{}{}^2\right)}}
\end{align}

\bibliographystyle{plainnat}
\bibliography{stdp_refs}

\end{document}